\documentclass[aps,prd,showpacs,superscriptaddress,nofootinbib,floatfix,showkeys,10pt]{revtex4-2}

\usepackage{bm}
\usepackage{times}
\usepackage{braket}
\usepackage{amsfonts,amssymb,latexsym,amsmath}
\usepackage[usenames,dvipsnames]{color}
\usepackage{graphicx}
\usepackage{slashed}
\usepackage{orcidlink}
\usepackage{multirow}
\usepackage{makecell}
\usepackage{booktabs}
\usepackage{array}
\usepackage{hyperref}

\renewcommand{\arraystretch}{1.6}
\allowdisplaybreaks[1]

\definecolor{nicered}{rgb}{0.7,0.1,0.1}
\definecolor{nicegreen}{rgb}{0.1,0.5,0.1}
\definecolor{navyblue}{RGB}{0, 110, 184}
\hypersetup{colorlinks,citecolor=nicegreen,linkcolor=nicered,urlcolor=navyblue}

\newcommand{\ii}{\mathrm{i}}
\newcommand{\dd}{\mathrm{d}}
\newcommand{\bsigma}{\boldsymbol{\sigma}}
\newcommand{\beps}{\boldsymbol{\epsilon}}
\newcommand{\bepsp}{\boldsymbol{\epsilon}'{}^*}
\newcommand{\hbk}{\hat{\bm{k}}}
\newcommand{\hbkp}{\hat{\bm{k}}'}
\newcommand{\bE}{\bm{E}}
\newcommand{\bH}{\bm{H}}
\newcommand{\bs}{\bm{s}}
\newcommand{\bsp}{\bm{s}'{}^*}
\newcommand{\bQ}{\mathcal Q}
\newcommand{\scrM}{\mathcal M}
\newcommand{\calJ}{\mathcal J}
\newcommand{\calG}{\mathcal G}

\begin{document}

\title{Spin-dependent polarizabilities of heavy vector mesons}

\author{Hao Dang\,\orcidlink{0000-0002-4320-2222}}
\email{haodang@stu.pku.edu.cn}
\affiliation{School of Physics, Peking University, Beijing 100871, China}

\author{Liang-Zhen Wen\,\orcidlink{0009-0006-8266-5840}}
\email{wenlzh\_hep-th@stu.pku.edu.cn}
\affiliation{School of Physics, Peking University, Beijing 100871, China}

\author{Yan-Ke Chen\,\orcidlink{0000-0002-9984-163X}}
\email{chenyanke@pku.edu.cn}
\affiliation{School of Physics and Center of High Energy Physics,
Peking University, Beijing 100871, China}

\author{Shi-Lin Zhu\,\orcidlink{0000-0002-4055-6906}}
\email{zhusl@pku.edu.cn}
\affiliation{School of Physics and Center of High Energy Physics,
Peking University, Beijing 100871, China}

\begin{abstract}
We investigate the spin-dependent electromagnetic polarizabilities of the heavy vector mesons \(D^*\) and \(B^*\) in heavy meson chiral perturbation theory up to \(\mathcal O(p^3)\). Using a twelve-element tensor basis for the real-photon Compton scattering on a spin-1 target, we determine two scalar, four vector and six tensor polarizabilities. We take the charm- and bottom-sector axial couplings from the measured \(D^*\) width and lattice-QCD calculations, respectively, and estimate the magnetic couplings in the nonrelativistic constituent-quark model. In the charm sector, the proximity of the charged \(D\pi\) thresholds generates strongly nonanalytic \(P\phi\)-loop contributions, producing large real contributions to several \(\bar D^{*0}\) polarizabilities and sizable imaginary parts for \(D^{*-}\), whose charged \(D\pi\) channel is open. The \(E1E1\)-type polarizabilities are enhanced much more strongly than their \(M1M1\)-type counterparts, consistent with the velocity suppression of the pion-cloud magnetic coupling. No analogous enhancement occurs for \(B^*\): the Born terms govern the magnetic polarizabilities, and the anomaly poles dominate the vector polarizabilities that receive no Born contribution. These results resolve the spin dependence of the \(D\pi\) threshold effect and provide benchmarks for future lattice-QCD studies.
\end{abstract}

\maketitle

\section{Introduction}
Understanding how the spectrum and internal structure of hadrons emerge from QCD remains a central goal of strong-interaction physics. The \(D^{(*)}\) and \(B^{(*)}\) systems provide a particularly instructive setting to address this question. Their simple valence composition, consisting of a heavy antiquark and a light quark, brings together short-distance dynamics associated with the heavy-quark mass and nonperturbative dynamics governing the light degrees of freedom. The interplay between these regimes shapes the spectrum and leads to nontrivial manifestations of approximate heavy-quark and chiral symmetries. In the heavy-quark limit, the heavy antiquark acts as a static color source, while chiral interactions with Goldstone bosons probe the light degrees of freedom. These features make heavy mesons a relatively clean setting to investigate the low-energy flavor and spin dynamics associated with the light quark, while comparisons between the charm and bottom sectors provide useful assessments of departures from the heavy-quark-limit picture~\cite{Wise:1992hn,Burdman:1992gh,Yan:1992gz}.

Electromagnetic observables provide complementary probes of this interplay and offer a detailed view of the resulting hadron structure. These include electromagnetic form factors and charge radii~\cite{Hwang:2001zd,Hwang:2001th,Luan:2015goa,Li:2017eic,Li:2019DsStar,Cui:2023DStar,Ahmed:2023zkk,Xu:2024fun,Hernandez-Pinto:2024kwg,Miramontes:2025vzb,Mutke:2026svq}, magnetic moments~\cite{Bose:1980vy,Becirevic:2009xp,Luan:2015goa,Simonis:2016pnh,Priyadarsini:2016tiu,Wang:2019mhm,Aliev:2019lsd,Xu:2024fun,Hernandez-Pinto:2024kwg}, and higher-order response coefficients. Electromagnetic polarizabilities constitute an important class of such coefficients, characterizing the response of a hadron's internal charge and magnetization distributions to external electromagnetic fields or, equivalently, parametrizing the structure-dependent terms in the low-energy Compton scattering amplitude~\cite{Holstein:2013kia,Pasquini:2018wbl}. However, a direct experimental determination of these quantities is particularly challenging for heavy vector mesons, since the \(D^{*}\) and \(B^{*}\) states decay too rapidly through strong or radiative processes to serve as conventional Compton scattering targets~\cite{ParticleDataGroup:2024cfk}. In the absence of such measurements, lattice QCD offers a complementary first-principles means of accessing the same structure-dependent electromagnetic response. Background-field methods for determining electromagnetic and spin polarizabilities, including those of charged targets and composite spin-1 systems, are well established and were applied in early studies of neutral vector mesons~\cite{Detmold:2006vu,Davoudi:2015zda,Christensen:2004ca,Detmold:2009dx}. Position-space Compton tensors have enabled a calculation of the pion electric polarizability at the physical pion mass~\cite{Feng:2022rkr}, while four-point correlation functions have been tested for charged-pion electric and magnetic polarizabilities in quenched calculations with heavier pion masses~\cite{Lee:2023rmz,Lee:2023lnx}. Spin-dependent magnetic responses and tensor polarizabilities of light vector mesons have also been investigated in background-field lattice studies~\cite{Luschevskaya:2018chr,Teryaev:2024men}. Recent lattice studies of nucleon electric polarizabilities, forward Compton amplitudes, and \(N\pi\) transition matrix elements have further emphasized the importance of explicitly resolving \(N\pi\) intermediate states~\cite{Wang:2023omf,Fu:2024gxq,Gao:2025loz}. At present, however, no lattice calculation of the heavy mesons polarizabilities has been reported. Effective field theory calculations therefore provide timely benchmarks and help identify the low-energy dynamics that future simulations must resolve.

Chiral effective field theory provides a systematic framework in which structure-dependent responses are organized order by order~\cite{Bernard:1991rq,Cho:1992nt,Bernard:1993bg,Hemmert:1996rw,Hemmert:1997tj,Gellas:2000mx,Bernard:2002pw}. Classic studies of nucleon polarizabilities established that chiral loops generate characteristic nonanalytic contributions and can substantially shape the low-energy electromagnetic response~\cite{Bernard:1991ru,Bernard:1993ry}. The application of chiral perturbation theory to massive matter fields requires special care because their masses remain finite in the chiral limit, and a naive relativistic expansion does not exhibit homogeneous chiral power counting. For heavy mesons, heavy meson chiral perturbation theory (HMChPT) provides an appropriate low-energy formulation by factoring out the large heavy-meson mass and introducing velocity-dependent fields that carry only residual momenta, while consistently implementing both chiral symmetry and heavy-quark spin symmetry~\cite{Casalbuoni:1996pg,Scherer:2002tk,Meng:2022ozq}. Recent chiral-EFT studies have extended scalar-polarizability calculations to spin-\(1/2\) and spin-\(3/2\) singly heavy baryons~\cite{Chen:2024xks,Wen:2025xed,Li:2026bmf}. In parallel, the scalar polarizabilities of heavy mesons and doubly heavy baryons have been analyzed in a unified heavy-hadron framework using heavy-diquark--antiquark symmetry~\cite{Dang:2026bxe}. In the charmed vector meson sector, the latter analysis identified an especially pronounced threshold effect: the mass splitting \(\Delta=m_{D^*}-m_D\) lies remarkably close to the pion mass. The resulting proximity to the \(D\pi\) threshold produces pronounced nonanalytic behavior in the chiral loop functions, strongly enhancing certain \(D^*\) polarizabilities and generating imaginary parts when an intermediate channel becomes kinematically accessible. This enhancement reveals an unusually strong sensitivity of the \(D^*\) electromagnetic response to low-energy threshold dynamics.

For a spin-1 target, however, the electromagnetic response is richer. It contains several independent spin structures and therefore cannot be fully captured by a spin-averaged description. Resolving these structures is essential for determining how the near-threshold enhancement identified above is distributed across the full electromagnetic response. Bilinears of the initial and final spin-1 polarization vectors naturally separate into scalar, vector, and symmetric-traceless rank-two tensor components. Correspondingly, the low-energy Compton amplitude encodes not only the ordinary scalar electric and magnetic polarizabilities, but also four vector polarizabilities associated with the rank-one component and additional tensor polarizabilities specific to targets with spin \(s\geq1\). While the four vector polarizabilities of the nucleon have been studied extensively in chiral effective theory and through dispersion relations~\cite{Hemmert:1996rw,Hemmert:1997tj,VijayaKumar:2000pv,Ragusa:1993rm,Pasquini:2018wbl,Babusci:1998ww}, the full set of spin-dependent polarizabilities of heavy vector mesons, including both vector and tensor components, has not yet received a systematic treatment. Deriving these quantities is a natural and necessary step toward dissecting the near-threshold mechanisms at a fully spin-resolved level.

In this work, we present a complete calculation of the spin-dependent electromagnetic polarizabilities of the heavy vector mesons \(D^*\) and \(B^*\) in HMChPT up to \(\mathcal O(p^3)\). We construct the real-photon Compton scattering amplitude for a spin-1 target using a twelve-element tensor basis and match its forward low-energy expansion to the standard scalar, vector, and tensor polarizabilities. Particular attention is devoted to the aforementioned near-threshold singularity in the charmed sector. We systematically contrast the polarizabilities of the charmed sector with those of the bottom sector where the corresponding mass splitting is significantly smaller than the pion mass.

In Sec.~\ref{sec:theory}, we define the spin-1 Compton amplitude, the associated twelve-element basis, the polarizabilities, and the HMChPT Lagrangian used in the calculation. Section~\ref{sec:numerical} presents the numerical inputs and our main results. The loop integrals and the exact analytic expressions for the projected forward limits are collected in Appendices~\ref{app:loop_integrals} and~\ref{app:analytic}.

\section{Theoretical Framework}\label{sec:theory}

\subsection{Spin-dependent polarizabilities for a spin-1 target}

The low-energy electromagnetic response of a spin-1 target can be organized according to the irreducible scalar, vector, and symmetric-traceless tensor structures formed by the initial and final polarization vectors~\cite{Chen:1998vi,ji:2003ia,Chen:2004wwa}. The scalar and vector parts of the effective interaction follow the standard low-energy multipole convention~\cite{Ragusa:1993rm,Babusci:1998ww,Holstein:2013kia,Li:2026bmf}. For the rank-two response, we construct the tensor terms from the symmetric-traceless spin tensor and the independent electromagnetic multipoles through the order considered. We thus define the complete interaction used in this work as
\begin{equation}\label{eq:Classical_Hamiltonian}
H_{\rm int}=H_{\rm scal}+H_{\rm vec}+H_{\rm tens},
\end{equation}
where
\begin{align}
	H_{\rm scal}
	&=-\frac12\,4\pi\left[
	\alpha_E\,\bE^2+\beta_M\,\bH^2
	\right],\notag \\
	H_{\rm vec}
	&=-\frac12\,4\pi\Big[
	\gamma_{E1E1}\,\bsigma\cdot(\bE\times\dot{\bE})
	+\gamma_{M1M1}\,\bsigma\cdot(\bH\times\dot{\bH})
	\notag\\
	&\quad+2\gamma_{E1M2}\,\sigma^iE^jH_{ij}
	+2\gamma_{M1E2}\,\sigma^iH^jE_{ij}
	\Big],\notag \\
	H_{\rm tens}
	&=-\frac12\,4\pi\Big[
	a_{E1E1}\,\bQ_{ij}E_iE_j
	+a_{M1M1}\,\bQ_{ij}H_iH_j
	\notag\\
	&\quad+a_{E2E2}\,\bQ_{ij}E_{ik}E_{jk}
	+a_{M2M2}\,\bQ_{ij}H_{ik}H_{jk}
	\notag\\
	&\quad+a_{E1E3}\,\bQ_{ij}E_kE_{k,ij}
	+a_{M1M3}\,\bQ_{ij}H_kH_{k,ij}
	\Big].\label{eq:H_svt}
\end{align}

\(H_{\rm scal}\) is spin independent, parameterizing the macroscopic deformation of the charge and magnetization distributions. \(H_{\rm vec}\) incorporates structures linear in the spin operator, defining four distinct vector polarizabilities that first appear at \(\mathcal O(\omega^{3})\). The tensor Hamiltonian \(H_{\rm tens}\) defines the six rank-two response coefficients retained. Such tensor responses occur only for targets with spin \(s \ge 1\). With the spin-1 polarization vectors \(\bm{\xi}\) and \(\bm{\xi}'\), we define
\begin{align}
	\bsigma&=\bm{\xi}'{}^*\times\bm{\xi},\notag\\
	\bQ_{ij}
	&=\xi_i'{}^*\xi_j+\xi_j'{}^*\xi_i
	-\frac23\delta_{ij}\,\bm{\xi}'{}^*\cdot\bm{\xi}.
\end{align}
Here \(\bsigma\) and \(\bQ_{ij}\) denote the rank-one and rank-two polarization bilinears, respectively.\footnote{For the conventional spin-1 generators \((S_i)_{jk}=-\ii\epsilon_{ijk}\), one has \(\langle\xi'|S_i|\xi\rangle=-\ii\sigma_i\). With \(T_{ij}=\frac12\{S_i,S_j\}-\frac23\delta_{ij}\), one also has \(\langle\xi'|T_{ij}|\xi\rangle=-Q_{ij}/2\). Thus, expressions written directly in terms of \(S_i\) and \(T_{ij}\) differ from our polarization-bilinear convention by these phase and normalization factors. The \(\ii J_i\) and \(-[J_iJ_j+J_jJ_i-\frac43\delta_{ij}]\) convention of Ref.~\cite{Chen:2004wwa} reproduces the bilinears used here.} The tensor \(\bQ_{ij}\) is symmetric and traceless, projecting the response onto the rank-two spin-tensor component. The derivative operators appearing in the tensor and mixed electric--magnetic terms are
\begin{align}
E_{ij}&=\frac12(\nabla_iE_j+\nabla_jE_i),&
H_{ij}&=\frac12(\nabla_iH_j+\nabla_jH_i),
\notag\\
E_{k,ij}&=\nabla_i\nabla_jE_k,&
H_{k,ij}&=\nabla_i\nabla_jH_k .
\end{align}

We perform the calculation in the target rest frame and adopt the Coulomb gauge. The incoming and outgoing photons have four-momenta \(k^\mu\) and \(k^{\prime\mu}\) and polarization four-vectors \(\epsilon^\mu\) and \(\epsilon^{\prime\mu}\), respectively, while the target four-velocity is \(v^\mu = (1, \mathbf{0})\). For convenience, we explicitly separate the temporal and three-dimensional spatial components:
\begin{equation}
	k^\mu = \omega(1, \hat{\boldsymbol{k}}), \quad k^{\prime\mu} = \omega(1, \hat{\boldsymbol{k}}'), \quad \epsilon^\mu = (0, \boldsymbol{\epsilon}), \quad \epsilon^{\prime\mu*} = (0, \boldsymbol{\epsilon}^{\prime*}),
\end{equation}
where \(\omega\) is the photon energy, and \(\hat{\boldsymbol{k}}\) and \(\hat{\boldsymbol{k}}'\) are unit vectors along the incoming and outgoing photon momenta, respectively (\(|\hat{\boldsymbol{k}}| = |\hat{\boldsymbol{k}}'| = 1\)). The scattering angle \(\theta\) is defined by the inner product \(\hat{\boldsymbol{k}} \cdot \hat{\boldsymbol{k}}' = \cos\theta\). In the Coulomb gauge (\(v \cdot \epsilon = v \cdot \epsilon' = 0\)), the transversality conditions are
\begin{equation}
	\boldsymbol{\epsilon} \cdot \hat{\boldsymbol{k}} = 0, \quad \boldsymbol{\epsilon}^{\prime*} \cdot \hat{\boldsymbol{k}}' = 0.
\end{equation}
The corresponding magnetic polarization vectors are defined as
\begin{equation}
	\bm{s}=\hat{\bm{k}}\times\bm{\epsilon},\qquad
	\bm{s}'{}^*=\hat{\bm{k}}'\times\bm{\epsilon}'{}^* .
\end{equation}

With the convention \(\scrM=\langle f|-H_{\rm int}|i\rangle\), the amplitude is systematically expanded in a complete basis as
\begin{equation}
\scrM=\sum_{i=1}^{12}A_i(\omega,\theta)\rho_i .
\label{eq:rho_basis_sum}
\end{equation}
The scalar functions \(A_i(\omega,\theta)\) encapsulate the dynamical structure of the target, while the \(\rho_i\) enumerate a complete set of independent tensor structures formed from the spin and polarization vectors. A convenient choice satisfying parity, time-reversal invariance, and real-photon transversality is~\cite{Chen:2004wwa}
\begin{align}
\rho_1&=\bepsp\!\cdot\beps\,
\bm{\xi}'{}^*\!\cdot\bm{\xi},
\notag\\
\rho_2&=\bsp\!\cdot\bs\,
\bm{\xi}'{}^*\!\cdot\bm{\xi},
\notag\\
\rho_3&=(\bm{\xi}'{}^*\times\bm{\xi})\!\cdot
(\bepsp\times\beps),
\notag\\
\rho_4&=(\bm{\xi}'{}^*\times\bm{\xi})\!\cdot
(\bsp\times\bs),
\notag\\
\rho_5&=(\bm{\xi}'{}^*\times\bm{\xi})\!\cdot\hbk\,
\bsp\!\cdot\beps
-(\bm{\xi}'{}^*\times\bm{\xi})\!\cdot\hbkp\,
\bepsp\!\cdot\bs,
\notag\\
\rho_6&=(\bm{\xi}'{}^*\times\bm{\xi})\!\cdot\hbkp\,
\bsp\!\cdot\beps-(\bm{\xi}'{}^*\times\bm{\xi})\!\cdot\hbk\,
\bepsp\!\cdot\bs,
\notag\\
\rho_7&=\bm{\xi}'{}^*\!\cdot\bepsp\,
\bm{\xi}\!\cdot\beps
+\bm{\xi}'{}^*\!\cdot\beps\,
\bm{\xi}\!\cdot\bepsp
-\frac23\,\bm{\xi}'{}^*\!\cdot\bm{\xi}\,
\bepsp\!\cdot\beps,
\notag\\
\rho_8&=\bm{\xi}'{}^*\!\cdot\bsp\,
\bm{\xi}\!\cdot\bs
+\bm{\xi}'{}^*\!\cdot\bs\,
\bm{\xi}\!\cdot\bsp
-\frac23\,\bm{\xi}'{}^*\!\cdot\bm{\xi}\,
\bsp\!\cdot\bs,
\notag\\
\rho_9&=\bepsp\!\cdot\hbk\,
\left(\bm{\xi}'{}^*\!\cdot\hbkp\,\bm{\xi}\!\cdot\beps
+\bm{\xi}'{}^*\!\cdot\beps\,\bm{\xi}\!\cdot\hbkp\right)
\notag\\
&\quad
+\beps\!\cdot\hbkp\,
\left(\bm{\xi}'{}^*\!\cdot\hbk\,\bm{\xi}\!\cdot\bepsp
+\bm{\xi}'{}^*\!\cdot\bepsp\,\bm{\xi}\!\cdot\hbk\right)
-\frac43\,\bm{\xi}'{}^*\!\cdot\bm{\xi}\,
\bepsp\!\cdot\hbk\,\beps\!\cdot\hbkp,
\notag\\
\rho_{10}&=\bsp\!\cdot\hbk\,
\left(\bm{\xi}'{}^*\!\cdot\hbkp\,\bm{\xi}\!\cdot\bs
+\bm{\xi}'{}^*\!\cdot\bs\,\bm{\xi}\!\cdot\hbkp\right)
\notag\\
&\quad
+\bs\!\cdot\hbkp\,
\left(\bm{\xi}'{}^*\!\cdot\hbk\,\bm{\xi}\!\cdot\bsp
+\bm{\xi}'{}^*\!\cdot\bsp\,\bm{\xi}\!\cdot\hbk\right)
-\frac43\,\bm{\xi}'{}^*\!\cdot\bm{\xi}\,
\bsp\!\cdot\hbk\,\bs\!\cdot\hbkp,
\notag\\
\rho_{11}&=\bepsp\!\cdot\beps
\left[
\bm{\xi}'{}^*\!\cdot\hbk\,\bm{\xi}\!\cdot\hbk
+\bm{\xi}'{}^*\!\cdot\hbkp\,\bm{\xi}\!\cdot\hbkp
-\frac23\,\bm{\xi}'{}^*\!\cdot\bm{\xi}
\right],
\notag\\
\rho_{12}&=\bsp\!\cdot\bs
\left[
\bm{\xi}'{}^*\!\cdot\hbk\,\bm{\xi}\!\cdot\hbk
+\bm{\xi}'{}^*\!\cdot\hbkp\,\bm{\xi}\!\cdot\hbkp
-\frac23\,\bm{\xi}'{}^*\!\cdot\bm{\xi}
\right].
\label{eq:rho}
\end{align}
Here the basis separates the low-energy response into three blocks. The structures \(\rho_{1,2}\) are spin independent, \(\rho_{3-6}\) are linear in the spin vector, and \(\rho_{7-12}\) contain the symmetric-traceless tensor component. In the calculation, we first evaluate the amplitude at nonzero scattering angle and project it onto the twelve independent structures. We take the limit \(\theta\to0\) after extracting the coefficient functions \(A_i\). If forward kinematics were imposed before this projection, \(\rho_9\) and \(\rho_{10}\) would vanish and several other structures would become linearly dependent, so that \(A_9\), \(A_{10}\), and some of the remaining amplitudes could not be separated. In what follows, \(A_i(\omega,0)\) denotes this ordered forward limit. The polarizabilities are therefore extracted from the forward limits of the amplitudes,
\begin{equation}
\Pi_i=
\left.\frac{1}{4\pi n_i !}\frac{\partial^{n_i}}{\partial \omega^{n_i}}A_{i}(\omega,0)\right|_{\omega=0},
\qquad
n_i=
\begin{cases}
2, & i=1,2,7,8,\\
3, & i=3,4,5,6,\\
4, & i=9,10,11,12.
\end{cases}.
\label{eq:Pi_definition}
\end{equation}
The derivative orders \(n_{i}\) are dictated by the multipole hierarchy of the corresponding operators: scalar and leading tensor polarizabilities enter at \(\mathcal O(\omega^{2})\), vector polarizabilities at \(\mathcal O(\omega^{3})\), and subleading tensor responses at \(\mathcal O(\omega^{4})\). Matching the low-energy expansion of Eq.~\eqref{eq:rho_basis_sum} to the Hamiltonian in Eq.~\eqref{eq:Classical_Hamiltonian} yields the following matching relations:
\begin{align}
\alpha_{E}&=\Pi_1,&
\beta_{M}&=\Pi_2,
\notag\\
\gamma_{E1E1}&=-(\Pi_3+\Pi_6),&
\gamma_{M1M1}&=-(\Pi_4+\Pi_5),
\notag\\
\gamma_{M1E2}&=\Pi_5,&
\gamma_{E1M2}&=-\Pi_6,
\notag\\
a_{E1E1}&=\Pi_7,&
a_{M1M1}&=\Pi_8,
\notag\\
a_{E2E2}&=2\Pi_9,&
a_{M2M2}&=2\Pi_{10},
\notag\\
a_{E1E3}&=-\Pi_{11},&
a_{M1M3}&=-\Pi_{12}.
\label{eq:rho_to_tensor}
\end{align}
Equation~\eqref{eq:rho_to_tensor} establishes the exact correspondence between the static polarizabilities and the multipole basis. \(\Pi_{1,2}\) reproduce the ordinary scalar electric and magnetic polarizabilities, \(\Pi_{3-6}\) determine the four vector polarizabilities, and \(\Pi_{7-12}\) define the six tensor polarizabilities of the spin-1 target. Since the \(\Pi_i\) have already been defined as derivatives at \(\omega=0\), the static polarizabilities in Eq.~\eqref{eq:rho_to_tensor} carry no residual photon-energy dependence.

\subsection{Heavy meson chiral Lagrangian}
We employ standard heavy meson chiral perturbation theory (HMChPT) conventions~\cite{Wise:1992hn,Burdman:1992gh,Yan:1992gz,Casalbuoni:1996pg}, with the field normalization and flavor assignments chosen as in Ref.~\cite{Dang:2026bxe}. In the heavy-hadron formalism, the common heavy-quark mass scale is explicitly factored out from the hadron fields, ensuring that the effective fields carry only small residual momenta. The pseudo-Goldstone bosons associated with the spontaneously broken chiral \(SU(3)_L \times SU(3)_R\) symmetry are parameterized by the nonlinear realization:
\begin{equation}
U=u^2=e^{i\phi/F_\phi},
\end{equation}
where \(F_{\phi}\) is the Goldstone decay constant, and we adopt \(F_\pi=92.4\) MeV, \(F_K=113\) MeV, and \(F_\eta=116\) MeV. The octet meson matrix is given by:
\begin{equation}
\phi=
\begin{pmatrix}
\pi^{0}+\frac{1}{\sqrt{3}}\eta & \sqrt{2}\pi^{+} & \sqrt{2}K^{+} \\
\sqrt{2}\pi^{-} & -\pi^{0}+\frac{1}{\sqrt{3}}\eta & \sqrt{2}K^{0} \\
\sqrt{2}K^{-} & \sqrt{2}\bar{K}^{0} & -\frac{2}{\sqrt{3}}\eta
\end{pmatrix}.
\end{equation}
The heavy mesons are organized into \(SU(3)_f\) flavor triplets. For the charmed sector, the pseudoscalar and vector heavy-meson triplets are defined as:
\begin{equation}
P=(\bar D^0,D^-,D_s^-),\qquad
P_\mu^*=(\bar D_\mu^{*0},D_\mu^{*-},D_{s\mu}^{*-}).
\end{equation}
The corresponding bottom multiplets are
\begin{equation}
P=(B^+,B^0,B_s^0),\qquad
P_\mu^*=(B_\mu^{*+},B_\mu^{*0},B_{s\mu}^{*0}) .
\end{equation}
Under the local \(SU(3)_f\) chiral transformation, the interactions between heavy mesons and Goldstone bosons are mediated by the chiral connection \(\Gamma_{\mu}\) and the axial-vector current \(u_{\mu}\):
\begin{align}
\Gamma_{\mu}&=\frac{1}{2}\left[u^\dagger\left(\partial_\mu-ir_\mu\right)u
+u\left(\partial_\mu-il_\mu\right)u^\dagger\right],
\notag\\
u_{\mu}&=\frac{\ii}{2}\left[u^{\dagger}\left(\partial_{\mu}-ir_{\mu}\right)u
-u\left(\partial_{\mu}-il_{\mu}\right)u^{\dagger}\right].
\end{align}
External electromagnetic fields are introduced by setting the right- and left-handed sources to \(r_{\mu} = l_{\mu} = -e Q A_{\mu}\). For the light Goldstone fields, the corresponding charge matrix is \(Q = Q_l = \mathrm{diag}(2/3, -1/3, -1/3)\). For the charmed and bottom heavy-meson triplets, the charge matrices \(Q\) are:
\begin{equation}
Q_c=\mathrm{diag}(0,-1,-1),\qquad
Q_b=\mathrm{diag}(1,0,0).
\end{equation}
The covariant derivatives acting on the pseudoscalar heavy-meson fields are consequently given by \(D_{\mu} P = \partial_{\mu} P + \Gamma_{\mu} P\), and similarly \(D_{\mu} P_{\nu}^* = \partial_{\mu} P_{\nu}^* + \Gamma_{\mu} P_{\nu}^*\) for the vector fields. Furthermore, the chiral electromagnetic field strengths are constructed as
\begin{align}
F_{\mu\nu}^{\pm}
&=u^\dagger F_{\mu\nu}^Ru\pm uF_{\mu\nu}^Lu^\dagger,
\notag\\
F_{\mu\nu}^{R,L}
&=\partial_\mu r_\nu-\partial_\nu r_\mu
-\ii\left[r_\mu,r_\nu\right],
\end{align}
To isolate the magnetic couplings in the effective Lagrangian, we employ the traceless combination:
\begin{equation}
\tilde F_{\mu\nu}^{\pm}
=F_{\mu\nu}^{\pm}-\frac13\operatorname{Tr}(F_{\mu\nu}^{\pm}) .
\end{equation}
The traceless part of the field strength tensor \(\tilde{F}_{\mu \nu}^{ \pm}\) is related to the traceless charge matrix of the light Goldstone fields \(Q_l\). The trace part \(\operatorname{Tr}\left(F_{\mu \nu}^{ \pm}\right)\) is related to the charge matrix of the heavy mesons \(Q_{c,b}\).

The leading-order (LO, \(\mathcal O(p^1)\)) heavy-meson Lagrangian governing the dynamics and interactions with Goldstone bosons is given by~\cite{Yan:1992gz,Casalbuoni:1996pg,Dang:2026bxe}:
\begin{equation}
\begin{aligned}
\mathcal{L}_{H\phi}^{(1)}
=&\,2\ii P^\dagger v\cdot D P
-2P^{*\dagger}(iv\cdot D-\Delta)P^{*}
\\
&+2g\left(\ii P^{*\dagger}_\mu u^\mu P+\mathrm{H.c.}\right)
-2\tilde g\,\ii\epsilon^{\mu \nu \rho \sigma}
u_\mu P^{*\dagger}_\nu P^{*}_\rho v_\sigma .
\end{aligned}
\label{eq:LO_HMChPT}
\end{equation}
Here \(\Delta=m_{P^*}-m_P\) is the hyperfine splitting. Heavy-quark spin symmetry relates the axial couplings of the \(P^*P\phi\) and \(P^*P^*\phi\) vertices, which yields \(\tilde{g}=g\). We impose this relation from the outset and denote both couplings by \(g\) in the following. As required by parity and angular momentum conservation, the \(PP\phi\) vertex is absent at this order. To incorporate the magnetic-dipole transitions and the \(\mathcal O(1/m_Q)\) recoil corrections necessary for Compton scattering, we include the next-to-leading order (NLO, \(\mathcal O(p^2)\)) electromagnetic Lagrangian~\cite{Amundson:1992yp,Cheng:1992xi,Wang:2019mhm}
\begin{align}
\mathcal{L}_{H\gamma}^{(2)}
=&-P^{\dagger}\frac{D^{2}}{m_P}P
+P^{*\dagger}\frac{D^{2}}{m_{P^*}}P^{*}
+4\ii\tilde{a}P^{*\dagger\mu}P^{*\nu}\tilde{F}_{\mu\nu}^+
+4\ii aP^{*\dagger\mu}P^{*\nu}\operatorname{Tr}(F_{\mu\nu}^+)
\notag\\
&+2\ii\tilde{a}\epsilon^{\mu\nu\rho\sigma}
P^{*\dagger}_{\rho}Pv_\sigma\tilde{F}_{\mu\nu}^+
+\mathrm{H.c.}
+2\ii a\epsilon^{\mu\nu\rho\sigma}
P^{*\dagger}_{\rho}Pv_\sigma\operatorname{Tr}(F_{\mu\nu}^+)
+\mathrm{H.c.}.
\label{eq:NLO_Hgamma}
\end{align}
The low-energy constants \(\tilde a\) and \(a\) parameterize the traceless light-flavor and flavor-singlet magnetic couplings, respectively. The chiral anomaly couples neutral Goldstone bosons to two photons through the Wess--Zumino--Witten functional~\cite{Wess:1971yu,Witten:1983tw}. The term relevant to real-photon Compton scattering is
\begin{equation}
	\mathcal{L}_{\phi\gamma\gamma}^{(4)}
	=
	-\frac{e^{2}}{32\pi^{2}}\,
	\epsilon^{\mu\nu\alpha\beta}
	F_{\mu\nu}F_{\alpha\beta}
	\left(
	\frac{\pi^{0}}{F_{\pi}}
	+\frac{\eta}{\sqrt{3}F_{\eta}}
	\right).
	\label{eq:WZW_phigammagamma}
\end{equation}
This neutral-Goldstone-boson pole mechanism originates from the chiral anomaly and is hereafter referred to as the anomaly pole contribution. It is the familiar pseudoscalar-pole contribution to spin-dependent Compton scattering~\cite{Hemmert:1997tj,VijayaKumar:2000pv}.

\begin{figure}[t]
\centering
\includegraphics[width=0.92\linewidth]{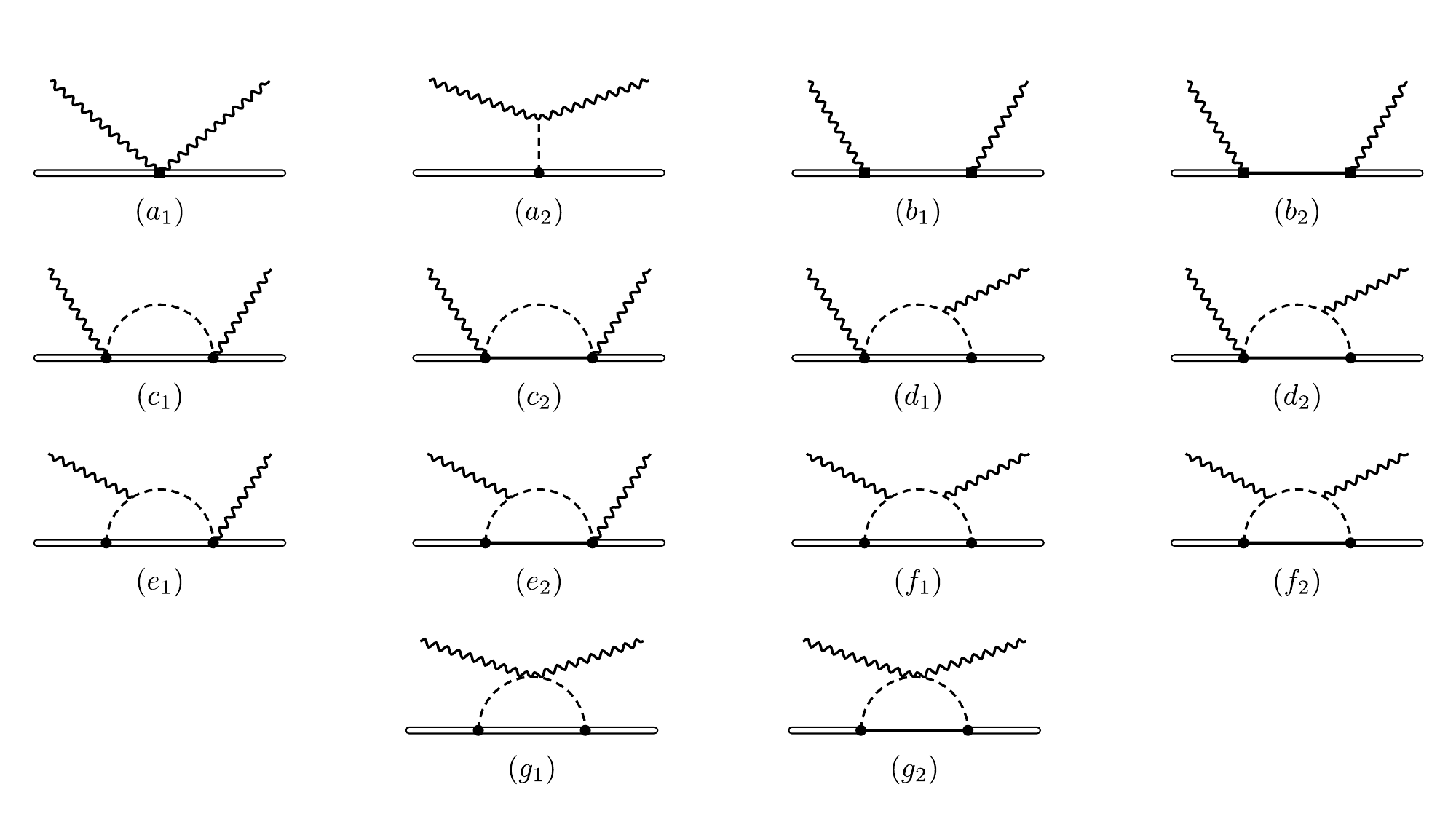}
\caption{Feynman diagrams contributing to the Compton
	amplitudes of heavy vector mesons up to \(\mathcal O(p^3)\).
	Double and single solid lines denote vector (\(P^*\)) and pseudoscalar
	(\(P\)) heavy mesons, respectively, while dashed and wavy lines denote
	Goldstone bosons and photons. Filled circles and squares represent
	vertices from \(\mathcal{L}_{H\phi}^{(1)}\) and
	\(\mathcal{L}_{H\gamma}^{(2)}\), respectively. Crossed diagrams are not
	shown. \((a_1)\), \((b_1)\), and \((b_2)\) are the Born
	diagrams, while \((a_2)\) is the anomaly pole
	diagram. \((c_1)\)--\((g_1)\) and
	\((c_2)\)--\((g_2)\) are loop diagrams with intermediate vector and
	pseudoscalar heavy mesons, respectively.}
\label{fig:spin1_loop_diagrams}
\end{figure}

For the systematic evaluation of loop diagrams, we adopt the standard chiral power counting scheme. The chiral dimension \(D_\chi\) of a connected diagram is given by~\cite{Bernard:1995dp,Scherer:2002tk}
\begin{equation}
D_\chi
=2L+1+\sum_d(d-2)N_d^\phi+\sum_d(d-1)N_d^{H\phi},
\label{eq:chiral_power_counting}
\end{equation}
where \(L\) represents the number of loops, while \(N_d^\phi\) and \(N_d^{H\phi}\) denote the number of vertices originating from the purely Goldstone and heavy-meson Lagrangians of chiral order \(d\), respectively. Applying this counting, we evaluate the scalar and spin-dependent polarizabilities through \(\mathcal O(p^3)\) from the Born, pole, and loop topologies displayed in Fig.~\ref{fig:spin1_loop_diagrams}.

Throughout this work, ``Born'' labels the tree-level heavy-meson topologies shown in Fig.~\ref{fig:spin1_loop_diagrams}. The pointlike Thomson amplitude generated by diagram \((a_1)\) is not included in the polarizabilities. All other nonvanishing contributions up to \(\mathcal O(p^3)\) are retained in the results. In particular, these include the magnetic-dipole \(P^*P\gamma\) transition pole from diagram \((b_2)\) and the anomaly pole from diagram \((a_2)\). Diagram \((b_1)\) does not contribute to the polarizabilities at this order. As low-energy checks, we verify that diagram \((a_1)\) reproduces the Thomson amplitude, and that the structure-dependent amplitudes entering the polarizability extraction vanish with the required powers of \(\omega\) as \(\omega\to0\). After all loop topologies and crossed diagrams are summed, the ultraviolet-divergent and renormalization-scale-dependent terms cancel. The resulting \(\mathcal O(p^3)\) polarizabilities are therefore ultraviolet finite and require no additional local counterterm at this order.

\section{Numerical Results}\label{sec:numerical}
The loop integrals and the analytic expressions for the amplitude functions \(A_i(\omega,0)\), defined as the ordered forward limits above, are collected in Appendices~\ref{app:loop_integrals} and~\ref{app:analytic}. To evaluate the polarizabilities numerically, we need to determine the axial coupling \(g\) in \(\mathcal{L}_{H\phi}^{(1)}\) and the magnetic low-energy constants \(a\) and \(\tilde a\) in \(\mathcal{L}_{H\gamma}^{(2)}\). We follow the numerical input conventions established in our previous work~\cite{Dang:2026bxe}. In the charm sector \(g\) is extracted from the measured strong decay width of \(D^{*+}\)~\cite{CLEO:2001foe},
\begin{equation}
\Gamma(D^{*+}\to D^0\pi^+)
=\frac{g_c^2}{12\pi F_\pi^2}\,|\bm{p}_{\pi^+}|^3,
\label{eq:Dstar_width_input}
\end{equation}
where \(\bm{p}_{\pi^+}\) is the pion three-momentum in the rest frame of the decaying \(D^{*+}\). For bottom mesons, the strong decay \(B^*\to B\pi\) is kinematically closed, and we take the corresponding axial coupling from unquenched lattice QCD in the static-heavy-quark formulation~\cite{Ohki:2008py}. This input is consistent with a more precise static-limit determination and with a calculation performed directly at the physical bottom-quark mass~\cite{Bernardoni:2014kla,Flynn:2015xna}. The values used in the numerical analysis are therefore
\begin{equation}
g_c=0.59\pm0.01\pm0.07,\qquad
g_b=0.516\pm0.005\pm0.052.
\label{eq:axial_coupling_inputs}
\end{equation}
Here the subscripts identify the flavor of the heavy antiquark. For \(g_c\), the two uncertainties are obtained by propagating the statistical and systematic errors of the measured \(D^{*+}\) width, respectively. For \(g_b\), the first uncertainty is statistical, while the second is obtained by combining in quadrature the systematic uncertainties associated with the chiral extrapolation, perturbative renormalization of the axial current, and finite-lattice-spacing effectss~\cite{Ohki:2008py}.

The NLO magnetic couplings \(a\) and \(\tilde a\) in \(\mathcal{L}_{H\gamma}^{(2)}\) determine the magnetic-dipole transitions and the associated Born contributions to the full Compton amplitude. They are estimated by matching the magnetic interactions of the light quark and the heavy antiquark to the nonrelativistic constituent-quark model~\cite{Wang:2019mhm}. With the normalization used in Eq.~\eqref{eq:NLO_Hgamma}, this gives
\begin{equation}
\tilde a=-\frac{1}{8m_q},\qquad
a=\frac{1}{24m_{\bar Q}},
\label{eq:magnetic_lec_inputs}
\end{equation}
where \(m_{q}\) and \(m_{\bar{Q}}\) represent the constituent masses of the light quark and the heavy antiquark, respectively. The corresponding \(P^*P\gamma\) transition coefficient is
\begin{equation}
C_P=Q_q\tilde a+3Q_{\bar Q}a,
\label{eq:CPstar_definition}
\end{equation}
where \(Q_q\) and \(Q_{\bar Q}\) are the electric charges of the light quark and the heavy antiquark, respectively. The adopted constituent masses, following the model input of Ref.~\cite{Wang:2019mhm}, are summarized in Table~\ref{tab:constituent_quark_masses}. To account for the model dependence of this matching procedure, we assign a conservative \(10\%\) uncertainty to these magnetic couplings.

Furthermore, the pronounced kinematical proximity between the \(D^{*}\) mass and the \(D\pi\) threshold renders the charmed vector mesons acutely sensitive to the precise values of the physical mass splittings. Consequently, we evaluate the mass splitting of each intermediate channel using the averaged physical masses compiled by the Particle Data Group (PDG)~\cite{ParticleDataGroup:2024cfk}. For an external vector meson \(P^*_{\rm ext}\) and an intermediate \(P_{\rm int}\phi\) state, we define
\begin{equation}
\delta_{\rm ch}=m_{P_{\rm int}}-m_{P^*_{\rm ext}},\qquad
\epsilon_{\rm th}=\delta_{\rm ch}+M_\phi.
\label{eq:channel_threshold_definitions}
\end{equation}
Thus, \(\epsilon_{\rm th}>0\) denotes a closed channel and \(\epsilon_{\rm th}<0\) an open channel. The resulting channel-dependent splittings are listed in Table~\ref{tab:charged_pion_thresholds}.

\begin{table*}[t]
\centering
\caption{Channel-dependent mass splittings (in units of MeV) used in the
charged-pion \(P\phi\) loops of the nonstrange \(D^*\) mesons. The splittings
are taken from the PDG averaged physical masses~\cite{ParticleDataGroup:2024cfk}.}
\label{tab:charged_pion_thresholds}
\renewcommand{\arraystretch}{1.2}
\setlength{\tabcolsep}{8pt}
\begin{tabular}{ccrrrr}
\hline\hline
External state & Intermediate state & \(\delta_{\rm ch}\) & \(M_\phi\) &
\(\epsilon_{\rm th}\) & Status \\
\hline
\(\bar D^{*0}\) & \(D^-\pi^+\) & \(-137.19\) & \(139.57039\) & \(+2.38\) & closed \\
\(D^{*-}\) & \(\bar D^0\pi^-\) & \(-145.42\) & \(139.57039\) & \(-5.85\) & open \\
\hline\hline
\end{tabular}
\end{table*}

The open status in Table~\ref{tab:charged_pion_thresholds} indicates that the intermediate particles can simultaneously be on shell, so the amplitudes and polarizabilities acquire imaginary parts. In a general treatment, the electromagnetic properties of an unstable particle are defined through the residues at its complex pole~\cite{Gegelia:2010nmt,Djukanovic:2013mka}. Although the \(D^*\) meson is unstable against the strong decay to \(D\pi\), its width is extremely narrow. We have verified that using a complex-mass prescription changes the numerical results only slightly. Accordingly, the present calculation uses real \(D^*\)-meson masses as an approximation.

\begin{table}[t]
	\centering
	\caption{The masses of the constituent quarks (in units of GeV).}
	\label{tab:constituent_quark_masses}
	\setlength{\tabcolsep}{4.5pt}
	\begin{tabular}{ccccc}
		\hline\hline
		\(m_u\) & \(m_d\) & \(m_s\) & \(m_c\) & \(m_b\) \\
		\hline
		\(0.336\) & \(0.336\) & \(0.540\) & \(1.66\) & \(4.73\) \\
		\hline\hline
	\end{tabular}
\end{table}

The numerical results for the scalar and spin-dependent polarizabilities of the charmed and bottom mesons are summarized in Tables~\ref{tab:D_meson_physical_polarizabilities} and~\ref{tab:B_meson_physical_polarizabilities}, respectively, from which several general features emerge. First, the scalar polarizabilities \(\alpha_E\) and \(\beta_M\) coincide with our previous work~\cite{Dang:2026bxe}. This agreement provides a useful check of the present spin-1 decomposition, since the spin-independent part is recovered from the same twelve-amplitude basis used to define the vector and tensor polarizabilities. Second, the Born contribution is highly selective: at \(\mathcal O(p^3)\) it contributes only to \(\beta_M\), \(\gamma_{M1M1}\), and \(a_{M1M1}\). This pattern follows from the structure of diagram \((b_2)\) in Fig.~\ref{fig:spin1_loop_diagrams}, which represents two consecutive magnetic-dipole \(P^*P\gamma\) transitions. Consequently, these three polarizabilities are particularly sensitive to the transition coefficient \(C_P\) and to the hyperfine splitting \(\Delta=m_{P^*}-m_P\). The anomaly pole contribution is similarly selective and enters only the four vector polarizabilities, \(\gamma_{E1E1}\), \(\gamma_{M1M1}\), \(\gamma_{M1E2}\), and \(\gamma_{E1M2}\). It arises from \(t\)-channel \(\pi^0\) and \(\eta\) exchange through the \(\phi\gamma\gamma\) vertex generated by the chiral anomaly in Eq.~\eqref{eq:WZW_phigammagamma}, together with the axial \(P^*P^*\phi\) coupling, and is absent from the scalar and tensor polarizabilities.

The \(\mathcal O(p^3)\) amplitudes also lead to several simple algebraic relations among the vector polarizabilities. For the loop contributions one finds
 \begin{equation}
 	\gamma_{M1M1}^{\rm loop}
 	=\gamma_{M1E2}^{\rm loop}
 	=-\gamma_{E1M2}^{\rm loop}.
 \end{equation}
These relations should be regarded as leading-order consequences of the restricted \(\mathcal O(p^3)\) operator and loop structure. Explicit \(\mathcal O(p^4)\) studies of nucleon spin polarizabilities~\cite{Gellas:2000mx,VijayaKumar:2000pv} show that subleading chiral contributions can induce deviations from analogous leading-order patterns.

For the charmed vector mesons, the dominant qualitative feature emerging from the \(P\phi\) loop channel is a pronounced near-threshold enhancement. This is the same near-threshold mechanism identified in the scalar-polarizability analysis of our previous work~\cite{Dang:2026bxe}: the \(D^*-D\) mass splitting \(\Delta\) lies remarkably close to the pion mass \(M_\pi\), causing the intermediate \(D\pi\) state to produce a pronounced nonanalytic dependence in the chiral loop functions. Consequently, the \(\bar{D}^{*0}\) channel exhibits exceptionally large real contributions, while the \(D^{*-}\) channel develops sizable imaginary parts once the threshold becomes kinematically open.

Crucially, the threshold enhancement is highly selective. As shown previously for the spin-averaged response~\cite{Dang:2026bxe}, it is concentrated primarily in \(\alpha_E\) rather than \(\beta_M\). The spin-resolved sector exhibits the same clear hierarchy between the \(E1E1\)- and \(M1M1\)-type responses: the singularity strongly amplifies \(\gamma_{E1E1}\) and \(a_{E1E1}\), making their magnitudes much larger than those of their \(M1M1\)-type counterparts, \(\gamma_{M1M1}\) and \(a_{M1M1}\).

The pronounced hierarchy between the \(E1E1\)- and \(M1M1\)-type polarizabilities originates predominantly from the \(P\phi\) loop channel. Its physical origin can be understood from the different couplings of the spatially extended pion cloud to electric and magnetic probes. In the long-wavelength limit, second-order perturbation theory gives~\cite{Ericson:1973dtc}
\begin{equation}
	\alpha_E
	=
	2\alpha_{\rm em}
	\sum_{n\neq i}
	\frac{|\langle n|D_z|i\rangle|^2}{E_n-E_i},
	\qquad
	\beta_M
	=
	2\alpha_{\rm em}
	\sum_{n\neq i}
	\frac{|\langle n|M_z|i\rangle|^2}{E_n-E_i}.
	\label{eq:classical_polar}
\end{equation}
To isolate the physical scaling relevant to the \(P\phi\) channel, we retain the intermediate states containing the pseudoscalar heavy meson and the charged pion. The dipole operators associated with the pion cloud can then be written classically as
\begin{equation}
	\bm D_\pi
	=
	\int \dd^3r\,
	\bm r\,\rho_\pi(\bm r),
	\qquad
	\bm M_\pi
	=
	\frac12\int \dd^3r\,
	\bm r\times\bm j_\pi(\bm r),
	\qquad
	\bm j_\pi(\bm r)
	\simeq
	\rho_\pi(\bm r)\bm v_\pi(\bm r).
	\label{eq:pion_cloud_dipoles}
\end{equation}
Here \(\bm D_\pi\) and \(\bm M_\pi\) denote only the contributions generated by the charged-pion cloud in the \(P\phi\) loop. The electric operator directly probes the spatial displacement of its charge distribution, whereas the magnetic operator contains the convective current and hence one explicit power of the characteristic pion velocity. Since the same near-threshold intermediate state supplies the energy denominator in both cases, their matrix elements scale parametrically as
\begin{equation}
	\frac{|\langle n|M_{\pi,z}|i\rangle|}
	{|\langle n|D_{\pi,z}|i\rangle|}
	\sim \mathcal O(v_\pi),
	\qquad
	\frac{\beta_M^{(P\phi)}}{\alpha_E^{(P\phi)}}
	\sim \mathcal O(v_\pi^2) .
\end{equation}
For the subthreshold \(D\pi\) channel, the characteristic momentum and velocity scales are
\begin{equation*}
 q_\pi\sim \sqrt{M_\pi^2-\Delta^2},\quad
 v_\pi\sim \frac{|q_\pi|}{M_\pi}\sim 0.17
\end{equation*}
Because \(\Delta \approx M_\pi\), the characteristic pion velocity is small. In the near-threshold expansion, the convective magnetic-dipole coupling carries two additional powers of \(v_\pi\) relative to the electric-dipole coupling. Consequently, the threshold singularity produces a much stronger enhancement of the \(E1E1\)-type loop responses, accounting for their much larger magnitudes relative to the corresponding \(M1M1\)-type quantities.

The dipole-based argument above should not, however, be extrapolated directly to the complete set of higher-multipole polarizabilities. For mixed and higher-multipole responses, including the \(\gamma_{E1M2}\), \(\gamma_{M1E2}\), and \(a_{M2M2}\) structures, the interplay among different multipole components leads to threshold patterns beyond the simple velocity hierarchy established for the \(E1E1\)- and \(M1M1\)-type responses. To present these patterns systematically, we summarize in Table~\ref{tab:Pphi_threshold_scaling} the leading powers of \(v_\pi\) extracted from the analytic \(P\phi\)-loop expressions for the \(\bar D^{*0}\) and \(D^{*-}\) channels.
\begin{table*}[t]
\centering
\caption{Leading near-threshold scaling of the \(P\phi\)-loop contributions in terms of \(v_\pi\). Common coupling and mass factors are not shown. Only polarizabilities within the same dimensional group can be compared directly.}
\label{tab:Pphi_threshold_scaling}
\renewcommand{\arraystretch}{1.25}
\setlength{\tabcolsep}{1.5pt}
\begin{tabular*}{\textwidth}{@{\extracolsep{\fill}}l |c c c c|c c c c|c c c c@{}}
\hline
\hline
Dimension
& \multicolumn{4}{c|}{\(\mathrm{fm}^3\)}
& \multicolumn{4}{c|}{\(\mathrm{fm}^4\)}
& \multicolumn{4}{c}{\(\mathrm{fm}^5\)} \\
\hline
Polarizabilities
& \(\alpha_E\) & \(\beta_M\) & \(a_{E1E1}\) & \(a_{M1M1}\)
& \(\gamma_{E1E1}\) & \(\gamma_{M1M1}\) & \(\gamma_{M1E2}\) & \(\gamma_{E1M2}\)
& \(a_{E2E2}\) & \(a_{M2M2}\) & \(a_{E1E3}\) & \(a_{M1M3}\) \\
\hline
Scaling
& \(v_\pi^{-3}\) & \(v_\pi^{-1}\) & \(v_\pi^{-3}\) & \(v_\pi^{-1}\)
& \(v_\pi^{-5}\) & \(v_\pi^{-3}\) & \(v_\pi^{-3}\) & \(v_\pi^{-3}\)
& \(v_\pi^{-5}\) & \(v_\pi^{-5}\) & \(v_\pi^{-5}\) & \(v_\pi^{-3}\) \\
\hline
\hline
\end{tabular*}
\end{table*}

The bottom sector displays a qualitatively different pattern. Since the \(B^*-B\) splitting lies below the pion mass, the \(B\pi\) intermediate channel is kinematically closed, and the \(P^*\phi\) and \(P\phi\) loops remain regular and of natural size. The hierarchy of the \(B^*\) polarizabilities is therefore largely governed by the selective Born and pole mechanisms discussed above. The Born term dominates \(\beta_M\), \(\gamma_{M1M1}\), and \(a_{M1M1}\) because of the magnetic transition coefficient \(C_P\) and the small \(B^*-B\) hyperfine splitting. Its larger magnitude for \(B^{*+}\) reflects the light-flavor dependence of \(C_P\). In \(\gamma_{M1M1}\), the anomaly pole term provides the next-largest contribution, reinforcing the Born term for \(B^{*+}\) but partially canceling it for \(B^{*0}\) and \(B_s^{*0}\). For the other three vector polarizabilities, \(\gamma_{E1E1}\), \(\gamma_{M1E2}\), and \(\gamma_{E1M2}\), the Born term vanishes and the anomaly pole contribution becomes dominant, with the loops providing channel-dependent corrections. The remaining polarizabilities are controlled by the regular chiral loops, for which the \(P^*\phi\) and \(P\phi\) contributions are generally comparable and may interfere either constructively or destructively. The loop contributions to \(B^{*+}\) and \(B^{*0}\) are consequently very similar, whereas those of \(B_s^{*0}\) are suppressed by the absence of pion loops.

\begin{table*}[t]
  \renewcommand{\arraystretch}{1.25}
  \setlength{\tabcolsep}{3.0pt}
  \centering
\caption{Scalar, vector, and tensor polarizabilities of the $D^*$ mesons obtained from Eq.~\eqref{eq:rho_to_tensor}. The entries for $\alpha_E$, $\beta_M$, $a_{E1E1}$, and $a_{M1M1}$ are in units of $10^{-4}\,\mathrm{fm}^3$; the four $\gamma$ polarizabilities are in units of $10^{-4}\,\mathrm{fm}^4$; and $a_{E2E2}$, $a_{M2M2}$, $a_{E1E3}$, and $a_{M1M3}$ are in units of $10^{-4}\,\mathrm{fm}^5$. The pole column contains the combined $\pi^0$- and $\eta$-exchange contributions. Numbers in parentheses denote uncertainties propagated from the axial and magnetic couplings, which are treated as independent. For the $D^{*-}$, the real and imaginary parts are displayed on successive lines.}
  \label{tab:D_meson_physical_polarizabilities}
  \begin{tabular*}{\textwidth}{@{\extracolsep{\fill}} c l c c c c c @{}}
    \toprule
    Polarizability & Meson & Born & $\pi^0,\eta$ poles & $P^*\phi$ loop & $P\phi$ loop & Total \\
    \midrule
    \multirow{4}{*}{$\alpha_E$}
    & $\bar D^{*0}$ & $0$ & $0$ & $2.15(52)$ & $291(70)$ & $294(70)$ \\
    & $D^{*-}$ & $0$ & $0$ & $1.81(43)$ & $-0.396(95)$ & $1.41(34)$ \\[-0.25ex]
    & {} & {} & {} & {} & $-64.4(154)\,\ii$ & $-64.4(154)\,\ii$ \\[0.50ex]
    & $D_s^{*-}$ & $0$ & $0$ & $0.341(82)$ & $0.395(95)$ & $0.736(176)$ \\
    \midrule
    \multirow{4}{*}{$\beta_M$}
    & $\bar D^{*0}$ & $-3.75(64)$ & $0$ & $0.215(52)$ & $0.979(235)$ & $-2.55(70)$ \\
    & $D^{*-}$ & $-0.229(83)$ & $0$ & $0.181(43)$ & $-0.0568(136)$ & $-0.105(88)$ \\[-0.25ex]
    & {} & {} & {} & {} & $+0.618(148)\,\ii$ & $+0.618(148)\,\ii$ \\[0.50ex]
    & $D_s^{*-}$ & $-0.0306(209)$ & $0$ & $0.0341(82)$ & $0.0259(62)$ & $0.0294(254)$ \\
    \midrule
    \multirow{4}{*}{$\gamma_{E1E1}$}
    & $\bar D^{*0}$ & $0$ & $5.38(64)$ & $-0.644(154)$ & $-1.29(31){\times}10^{4}$ & $-1.29(31){\times}10^{4}$ \\
    & $D^{*-}$ & $0$ & $-5.25(63)$ & $-0.611(146)$ & $-0.180(43)$ & $-6.04(82)$ \\[-0.25ex]
    & {} & {} & {} & {} & $-1.10(26){\times}10^{3}\,\ii$ & $-1.10(26){\times}10^{3}\,\ii$ \\[0.50ex]
    & $D_s^{*-}$ & $0$ & $-0.136(16)$ & $-0.0325(78)$ & $-0.147(35)$ & $-0.316(59)$ \\
    \midrule
    \multirow{4}{*}{$\gamma_{M1M1}$}
    & $\bar D^{*0}$ & $-7.81(132)$ & $-5.38(64)$ & $-0.129(31)$ & $-67.4(162)$ & $-80.8(169)$ \\
    & $D^{*-}$ & $-0.478(173)$ & $5.25(63)$ & $-0.122(29)$ & $-0.0390(93)$ & $4.61(61)$ \\[-0.25ex]
    & {} & {} & {} & {} & $+16.0(38)\,\ii$ & $+16.0(38)\,\ii$ \\[0.50ex]
    & $D_s^{*-}$ & $-0.0638(436)$ & $0.136(16)$ & $-0.00650(156)$ & $-0.0185(44)$ & $0.0477(448)$ \\
    \midrule
    \multirow{4}{*}{$\gamma_{M1E2}$}
    & $\bar D^{*0}$ & $0$ & $5.38(64)$ & $-0.129(31)$ & $-67.4(162)$ & $-62.2(156)$ \\
    & $D^{*-}$ & $0$ & $-5.25(63)$ & $-0.122(29)$ & $-0.0390(93)$ & $-5.41(67)$ \\[-0.25ex]
    & {} & {} & {} & {} & $+16.0(38)\,\ii$ & $+16.0(38)\,\ii$ \\[0.50ex]
    & $D_s^{*-}$ & $0$ & $-0.136(16)$ & $-0.00650(156)$ & $-0.0185(44)$ & $-0.161(22)$ \\
    \midrule
    \multirow{4}{*}{$\gamma_{E1M2}$}
    & $\bar D^{*0}$ & $0$ & $5.38(64)$ & $0.129(31)$ & $67.4(162)$ & $72.9(168)$ \\
    & $D^{*-}$ & $0$ & $-5.25(63)$ & $0.122(29)$ & $0.0390(93)$ & $-5.08(59)$ \\[-0.25ex]
    & {} & {} & {} & {} & $-16.0(38)\,\ii$ & $-16.0(38)\,\ii$ \\[0.50ex]
    & $D_s^{*-}$ & $0$ & $-0.136(16)$ & $0.00650(156)$ & $0.0185(44)$ & $-0.111(10)$ \\
    \bottomrule
  \end{tabular*}
\end{table*}

\begin{table*}[t]
  \renewcommand{\arraystretch}{1.25}
  \setlength{\tabcolsep}{3.0pt}
  \centering
  \noindent TABLE~\ref{tab:D_meson_physical_polarizabilities} (continued).\quad Scalar, vector, and tensor polarizabilities of the $D^*$ mesons.

  \vspace{0.8ex}
  \begin{tabular*}{\textwidth}{@{\extracolsep{\fill}} c l c c c c c @{}}
    \toprule
    Polarizability & Meson & Born & $\pi^0,\eta$ poles & $P^*\phi$ loop & $P\phi$ loop & Total \\
    \midrule
    \multirow{4}{*}{$a_{E1E1}$}
    & $\bar D^{*0}$ & $0$ & $0$ & $-0.645(155)$ & $287(69)$ & $287(69)$ \\
    & $D^{*-}$ & $0$ & $0$ & $-0.543(130)$ & $-0.168(40)$ & $-0.712(171)$ \\[-0.25ex]
    & {} & {} & {} & {} & $-66.8(160)\,\ii$ & $-66.8(160)\,\ii$ \\[0.50ex]
    & $D_s^{*-}$ & $0$ & $0$ & $-0.102(25)$ & $0.291(70)$ & $0.189(45)$ \\
    \midrule
    \multirow{4}{*}{$a_{M1M1}$}
    & $\bar D^{*0}$ & $-5.62(95)$ & $0$ & $0.323(77)$ & $-2.94(70)$ & $-8.23(114)$ \\
    & $D^{*-}$ & $-0.344(125)$ & $0$ & $0.272(65)$ & $0.170(41)$ & $0.098(164)$ \\[-0.25ex]
    & {} & {} & {} & {} & $-1.86(44)\,\ii$ & $-1.86(44)\,\ii$ \\[0.50ex]
    & $D_s^{*-}$ & $-0.0459(314)$ & $0$ & $0.0512(123)$ & $-0.0777(186)$ & $-0.0724(320)$ \\
    \midrule
    \multirow{4}{*}{$a_{E2E2}$}
    & $\bar D^{*0}$ & $0$ & $0$ & $-0.551(132)$ & $9.70(233){\times}10^{3}$ & $9.70(233){\times}10^{3}$ \\
    & $D^{*-}$ & $0$ & $0$ & $-0.543(130)$ & $-0.116(28)$ & $-0.658(158)$ \\[-0.25ex]
    & {} & {} & {} & {} & $+834(200)\,\ii$ & $+834(200)\,\ii$ \\[0.50ex]
    & $D_s^{*-}$ & $0$ & $0$ & $-0.00814(195)$ & $0.0329(79)$ & $0.0247(59)$ \\
    \midrule
    \multirow{4}{*}{$a_{M2M2}$}
    & $\bar D^{*0}$ & $0$ & $0$ & $0.220(53)$ & $1.11(27){\times}10^{3}$ & $1.11(27){\times}10^{3}$ \\
    & $D^{*-}$ & $0$ & $0$ & $0.217(52)$ & $0.104(25)$ & $0.321(77)$ \\[-0.25ex]
    & {} & {} & {} & {} & $+127(30)\,\ii$ & $+127(30)\,\ii$ \\[0.50ex]
    & $D_s^{*-}$ & $0$ & $0$ & $0.00325(78)$ & $-0.00615(147)$ & $-0.00290(69)$ \\
    \midrule
    \multirow{4}{*}{$a_{E1E3}$}
    & $\bar D^{*0}$ & $0$ & $0$ & $-0.220(53)$ & $3.64(87){\times}10^{3}$ & $3.64(87){\times}10^{3}$ \\
    & $D^{*-}$ & $0$ & $0$ & $-0.217(52)$ & $-0.0490(117)$ & $-0.266(64)$ \\[-0.25ex]
    & {} & {} & {} & {} & $+312(75)\,\ii$ & $+312(75)\,\ii$ \\[0.50ex]
    & $D_s^{*-}$ & $0$ & $0$ & $-0.00325(78)$ & $0.0128(31)$ & $0.00956(229)$ \\
    \midrule
    \multirow{4}{*}{$a_{M1M3}$}
    & $\bar D^{*0}$ & $0$ & $0$ & $0.110(26)$ & $-38.7(93)$ & $-38.6(92)$ \\
    & $D^{*-}$ & $0$ & $0$ & $0.109(26)$ & $0.0452(108)$ & $0.154(37)$ \\[-0.25ex]
    & {} & {} & {} & {} & $+8.66(208)\,\ii$ & $+8.66(208)\,\ii$ \\[0.50ex]
    & $D_s^{*-}$ & $0$ & $0$ & $0.00163(39)$ & $-0.00391(94)$ & $-0.00228(55)$ \\
    \bottomrule
  \end{tabular*}
\end{table*}

\begin{table*}[t]
  \renewcommand{\arraystretch}{1.30}
  \setlength{\tabcolsep}{2.5pt}
  \centering
\caption{Scalar, vector, and tensor polarizabilities of the $B^*$ mesons obtained from Eq.~\eqref{eq:rho_to_tensor}. The units, pole contribution, and uncertainty convention are the same as in Table~\ref{tab:D_meson_physical_polarizabilities}.}
  \label{tab:B_meson_physical_polarizabilities}
  \begin{tabular*}{\textwidth}{@{\extracolsep{\fill}} c l c c c c c @{}}
    \toprule
    Polarizability & Meson & Born & $\pi^0,\eta$ poles & $P^*\phi$ loop & $P\phi$ loop & Total \\
    \midrule
    \multirow{3}{*}{$\alpha_E$}
    & $B^{*+}$ & $0$ & $0$ & $1.65(33)$ & $1.24(25)$ & $2.88(58)$ \\
    & $B^{*0}$ & $0$ & $0$ & $1.38(28)$ & $1.12(23)$ & $2.50(51)$ \\
    & $B_s^{*0}$ & $0$ & $0$ & $0.261(53)$ & $0.194(39)$ & $0.455(92)$ \\
    \midrule
    \multirow{3}{*}{$\beta_M$}
    & $B^{*+}$ & $-7.60(158)$ & $0$ & $0.165(33)$ & $0.101(20)$ & $-7.34(158)$ \\
    & $B^{*0}$ & $-2.34(44)$ & $0$ & $0.138(28)$ & $0.0884(179)$ & $-2.12(44)$ \\
    & $B_s^{*0}$ & $-0.982(177)$ & $0$ & $0.0261(53)$ & $0.0160(32)$ & $-0.940(178)$ \\
    \midrule
    \multirow{3}{*}{$\gamma_{E1E1}$}
    & $B^{*+}$ & $0$ & $4.71(48)$ & $-0.492(100)$ & $-1.13(23)$ & $3.08(15)$ \\
    & $B^{*0}$ & $0$ & $-4.59(46)$ & $-0.467(95)$ & $-1.11(23)$ & $-6.17(78)$ \\
    & $B_s^{*0}$ & $0$ & $-0.119(12)$ & $-0.0248(50)$ & $-0.0508(103)$ & $-0.195(27)$ \\
    \midrule
    \multirow{3}{*}{$\gamma_{M1M1}$}
    & $B^{*+}$ & $-50.0(104)$ & $-4.71(48)$ & $-0.0984(199)$ & $-0.176(36)$ & $-55.0(104)$ \\
    & $B^{*0}$ & $-15.4(29)$ & $4.59(46)$ & $-0.0935(189)$ & $-0.172(35)$ & $-11.1(29)$ \\
    & $B_s^{*0}$ & $-6.46(117)$ & $0.119(12)$ & $-0.00497(101)$ & $-0.00821(166)$ & $-6.35(117)$ \\
    \midrule
    \multirow{3}{*}{$\gamma_{M1E2}$}
    & $B^{*+}$ & $0$ & $4.71(48)$ & $-0.0984(199)$ & $-0.176(36)$ & $4.43(42)$ \\
    & $B^{*0}$ & $0$ & $-4.59(46)$ & $-0.0935(189)$ & $-0.172(35)$ & $-4.85(52)$ \\
    & $B_s^{*0}$ & $0$ & $-0.119(12)$ & $-0.00497(101)$ & $-0.00821(166)$ & $-0.133(15)$ \\
    \midrule
    \multirow{3}{*}{$\gamma_{E1M2}$}
    & $B^{*+}$ & $0$ & $4.71(48)$ & $0.0984(199)$ & $0.176(36)$ & $4.98(53)$ \\
    & $B^{*0}$ & $0$ & $-4.59(46)$ & $0.0935(189)$ & $0.172(35)$ & $-4.32(41)$ \\
    & $B_s^{*0}$ & $0$ & $-0.119(12)$ & $0.00497(101)$ & $0.00821(166)$ & $-0.106(9)$ \\
    \midrule
    \multirow{3}{*}{$a_{E1E1}$}
    & $B^{*+}$ & $0$ & $0$ & $-0.494(100)$ & $0.834(169)$ & $0.341(69)$ \\
    & $B^{*0}$ & $0$ & $0$ & $-0.415(84)$ & $0.766(155)$ & $0.350(71)$ \\
    & $B_s^{*0}$ & $0$ & $0$ & $-0.0783(159)$ & $0.130(26)$ & $0.0515(104)$ \\
    \midrule
    \multirow{3}{*}{$a_{M1M1}$}
    & $B^{*+}$ & $-11.4(24)$ & $0$ & $0.247(50)$ & $-0.302(61)$ & $-11.5(24)$ \\
    & $B^{*0}$ & $-3.52(66)$ & $0$ & $0.208(42)$ & $-0.265(54)$ & $-3.57(66)$ \\
    & $B_s^{*0}$ & $-1.47(27)$ & $0$ & $0.0392(79)$ & $-0.0479(97)$ & $-1.48(27)$ \\
    \midrule
    \multirow{3}{*}{$a_{E2E2}$}
    & $B^{*+}$ & $0$ & $0$ & $-0.421(85)$ & $0.921(187)$ & $0.500(101)$ \\
    & $B^{*0}$ & $0$ & $0$ & $-0.415(84)$ & $0.916(185)$ & $0.501(101)$ \\
    & $B_s^{*0}$ & $0$ & $0$ & $-0.00622(126)$ & $0.0119(24)$ & $0.00568(115)$ \\
    \midrule
    \multirow{3}{*}{$a_{M2M2}$}
    & $B^{*+}$ & $0$ & $0$ & $0.169(34)$ & $-0.248(50)$ & $-0.0799(162)$ \\
    & $B^{*0}$ & $0$ & $0$ & $0.166(34)$ & $-0.246(50)$ & $-0.0801(162)$ \\
    & $B_s^{*0}$ & $0$ & $0$ & $0.00249(50)$ & $-0.00346(70)$ & $-0.000967(196)$ \\
    \midrule
    \multirow{3}{*}{$a_{E1E3}$}
    & $B^{*+}$ & $0$ & $0$ & $-0.169(34)$ & $0.363(73)$ & $0.194(39)$ \\
    & $B^{*0}$ & $0$ & $0$ & $-0.166(34)$ & $0.361(73)$ & $0.195(39)$ \\
    & $B_s^{*0}$ & $0$ & $0$ & $-0.00249(50)$ & $0.00470(95)$ & $0.00221(45)$ \\
    \midrule
    \multirow{3}{*}{$a_{M1M3}$}
    & $B^{*+}$ & $0$ & $0$ & $0.0843(171)$ & $-0.138(28)$ & $-0.0542(110)$ \\
    & $B^{*0}$ & $0$ & $0$ & $0.0830(168)$ & $-0.137(28)$ & $-0.0543(110)$ \\
    & $B_s^{*0}$ & $0$ & $0$ & $0.00124(25)$ & $-0.00188(38)$ & $-0.00064(13)$ \\
    \bottomrule
  \end{tabular*}
\end{table*}

\section{Summary}

In this work, we have systematically calculated the spin-dependent electromagnetic polarizabilities of the heavy vector mesons \(D^*\) and \(B^*\) in HMChPT up to \(\mathcal O(p^3)\). Using a twelve-element tensor basis for real-photon Compton scattering on a spin-1 target, we determine the four vector and six tensor polarizabilities. The two scalar responses are recovered in the same decomposition and agree with our previous calculation. The charm- and bottom-sector axial couplings are taken from the measured \(D^*\) decay width and lattice-QCD calculations, respectively. The magnetic low-energy constants are estimated using the nonrelativistic constituent-quark model.

The calculation includes Born diagrams, anomaly pole diagrams, and chiral loops with intermediate vector and pseudoscalar heavy mesons. At the order considered, the Born term enters only \(\beta_M\), \(\gamma_{M1M1}\), and \(a_{M1M1}\), whereas the anomaly pole term is restricted to the four vector polarizabilities. The remaining responses are generated by the chiral loops.

The most striking feature of the charm sector arises from the proximity of the charged \(D\pi\) thresholds. The resulting nonanalytic behavior of the \(P\phi\)-loop functions produces exceptionally large real contributions to several \(\bar D^{*0}\) polarizabilities and sizable imaginary parts for \(D^{*-}\) when the corresponding intermediate channel is kinematically open. The \(E1E1\)-type polarizabilities \(\alpha_E\), \(\gamma_{E1E1}\), and \(a_{E1E1}\) are enhanced much more strongly than their \(M1M1\)-type counterparts, consistent with the velocity suppression of the pion-cloud magnetic coupling. The mixed and higher-multipole responses exhibit a broader range of threshold powers, as summarized in Table~\ref{tab:Pphi_threshold_scaling}.

The bottom sector displays no analogous near-threshold singularity, and its chiral-loop contributions remain of natural size. Its magnetic-type responses are governed primarily by Born terms, which dominate \(\beta_M\), \(\gamma_{M1M1}\) and \(a_{M1M1}\), while the anomaly pole term provides the leading contribution to \(\gamma_{E1E1}\), \(\gamma_{M1E2}\) and \(\gamma_{E1M2}\). The large charm-sector polarizabilities also indicate the quantitative scope of the present expansion: the nearby threshold scale is counted together with the ordinary chiral momentum, so the largest values remain sensitive to the precise threshold separation. A dedicated power counting that separates these scales would allow a more precise treatment~\cite{Fleming:2007rp,Alhakami:2015uea}. Our analytical and numerical results provide benchmarks for such refinements and for future lattice-QCD studies of heavy vector mesons.

\begin{acknowledgments}
This project was supported by the National Natural Science Foundation of China (Grant No.~12475137). The computational resources were supported by
the High-performance Computing Platform of Peking University. 
\end{acknowledgments}

\clearpage
\begin{appendix}

\section{Loop Integrals}\label{app:loop_integrals}

We first collect the loop-integral conventions used in the calculation. The propagator denominators are combined with Feynman parameters according to
\begin{equation}
	\frac{1}{A_1 A_2 \cdots A_n}=\int_0^1 d x_1 \cdots d x_n \delta\left(\sum x_i-1\right) \frac{(n-1)!}{\left[x_1 A_1+x_2 A_2+\cdots x_n A_n\right]^n}.
\end{equation}
The divergent loop integrals are regularized dimensionally and expanded about four spacetime dimensions. The tensor integrals needed below are reduced to the scalar functions defined in Refs.~\cite{Bernard:1995dp,Scherer:2002tk,Hemmert:1996rw}:
\begin{equation}
\begin{aligned}
	\frac{1}{i} &\int \frac{d^d \ell}{(2 \pi)^d} \frac{\left\{1, \ell_\mu \ell_\nu, \ell_\mu \ell_\nu \ell_\alpha \ell_\beta\right\}}{(v \cdot \ell-\omega-i \epsilon)\left(M_{\phi}^2-\ell^2-i \epsilon\right)}  \\
&=\left\{J_0\left(\omega, M_{\phi}^2\right)\right.,\quad g_{\mu \nu} J_2\left(\omega, M_{\phi}^2\right)+v_\mu v_\nu J_3\left(\omega, M_{\phi}^2\right),\quad \left.\left(g_{\mu \nu} g_{\alpha \beta}+\text { perm. }\right) J_6\left(\omega, M_{\phi}^2\right)+\ldots\right\}.
\end{aligned}
\end{equation}
Here $M_\phi$ denotes the Goldstone-boson mass. All loop integrals can be expressed in terms of the basis function $J_0$:
\begin{equation}\label{eq:J0_J2_J6}
\begin{aligned}
	J_0\left(\omega, M_{\phi}^2\right) & =-4 L \omega+\frac{\omega}{8 \pi^2}\left(1-2 \ln \frac{M_\phi}{\mu}\right)-\frac{1}{4 \pi^2} \sqrt{M_{\phi}^2-\omega^2} \arccos \frac{-\omega}{M_\phi}+\mathcal{O}(d-4),\\
J_2\left(\omega, M_{\phi}^2\right) & =\frac{1}{d-1}\left[\left(M_{\phi}^2-\omega^2\right) J_0\left(\omega, M_{\phi}^2\right)-\omega \Delta_{\chi}\right], \\
J_6\left(\omega, M_{\phi}^2\right) & =\frac{1}{d+1}\left[\left(M_{\phi}^2-\omega^2\right) J_2\left(\omega, M_{\phi}^2\right)-\frac{M_{\phi}^2 \omega}{d} \Delta_{\chi}\right].
\end{aligned}
\end{equation}
In Eq.~\eqref{eq:J0_J2_J6} we have used
\begin{equation}
\begin{aligned}
\Delta_{\chi} & =2 M_{\phi}^2\left(L+\frac{1}{16 \pi^2} \ln \frac{M_\phi}{\mu}\right) +\mathcal{O}(d-4),\\
L & =\frac{\mu^{d-4}}{16 \pi^2}\left[\frac{1}{d-4}+\frac{1}{2}\left(\gamma_{\rm E}-1-\ln 4 \pi\right)\right],
\end{aligned}
\end{equation}
where \(\gamma_{\rm E}=0.577215\) is the Euler--Mascheroni constant and \(\mu\) is the scale introduced by dimensional regularization.

In the forward Compton amplitude, the combinations
\(J_i(-\omega-\delta)\) and \(J_i(\omega-\delta)\) occur together.  It is
therefore useful to define the even and odd combinations
\begin{align}
\mathcal{J}_i(\omega,\delta,M_{\phi}^2)
&=J_i(\omega-\delta,M_{\phi}^2)+J_i(-\omega-\delta,M_{\phi}^2),
\notag\\
\mathcal{G}_i(\omega,\delta,M_{\phi}^2)
&=J_i(\omega-\delta,M_{\phi}^2)-J_i(-\omega-\delta,M_{\phi}^2).
\end{align}
Their derivatives with respect to \(M_\phi^2\) are denoted by
\begin{equation}
\begin{aligned}
\mathcal{J}_i^{\prime}\left(\omega,\delta ,M_{\phi}^2\right) & =\frac{\partial}{\partial\left(M_{\phi}^2\right)} \mathcal{J}_i\left(\omega, \delta,M_{\phi}^2\right) \\
\mathcal{J}_i^{\prime \prime}\left(\omega, \delta,M_{\phi}^2\right) & =\frac{\partial^2}{\partial\left(M_{\phi}^2\right)^2} \mathcal{J}_i\left(\omega,\delta, M_{\phi}^2\right),
\end{aligned}
\end{equation}
and analogously
\begin{equation}
\begin{aligned}
\mathcal{G}_i^{\prime}\left(\omega,\delta ,M_{\phi}^2\right) & =\frac{\partial}{\partial\left(M_{\phi}^2\right)} \mathcal{G}_i\left(\omega, \delta,M_{\phi}^2\right) \\
\mathcal{G}_i^{\prime \prime}\left(\omega, \delta,M_{\phi}^2\right) & =\frac{\partial^2}{\partial\left(M_{\phi}^2\right)^2} \mathcal{G}_i\left(\omega,\delta, M_{\phi}^2\right).
\end{aligned}
\end{equation}

\section{Analytic Expressions for the Projected Forward Limits}\label{app:analytic}

We collect here the analytic contributions to the projected forward limits \(A_i\equiv A_i(\omega,0)\) entering the definitions in Eq.~\eqref{eq:Pi_definition}. As explained in Sec.~\ref{sec:theory}, each amplitude is first separated at nonzero scattering angle, and only then is the limit \(\theta\to0\) taken. Throughout this appendix, \(\phi\) denotes a Goldstone boson, and \(M_\phi\) and \(F_\phi\) denote its mass and decay constant, respectively.  Diagram \((a_1)\) generates the Thomson amplitude for a pointlike target and therefore does not contribute to the polarizabilities. The remaining topologies up to \(\mathcal O(p^3)\) are the Born diagrams \((b_1)\) and \((b_2)\), the anomaly pole diagram \((a_2)\), the \(P^*\phi\) loops \((c_1)\)--\((g_1)\), and the \(P\phi\) loops \((c_2)\)--\((g_2)\), as shown in Fig.~\ref{fig:spin1_loop_diagrams}.  The loop functions \(\calJ_i\) and \(\calG_i\), together with their derivatives, are defined in
Appendix~\ref{app:loop_integrals}.

\subsection{Born Contributions}

The magnetic couplings in \(\mathcal L_{H\gamma}^{(2)}\) generate nonzero Compton amplitudes from both Born diagrams \((b_1)\) and \((b_2)\).  After the low-energy expansion and matching to the static polarizabilities, however, diagram \((b_1)\) gives no contribution up to \(\mathcal O(p^3)\).  We therefore display only the contribution from diagram \((b_2)\):
\begin{align}
A_2^{(b_2)}
&=-\frac{32}{3} e^2 C_P^2
\frac{\omega^2\Delta}{\Delta^2-\omega^2},
\notag\\
A_4^{(b_2)}
&=16 e^2 C_P^2
\frac{\omega^3}{\Delta^2-\omega^2},
\notag\\
A_8^{(b_2)}
&=-16 e^2 C_P^2
\frac{\omega^2\Delta}{\Delta^2-\omega^2}.
\label{eq:born_b2_forward}
\end{align}
The transition coefficient \(C_P\), defined in Eq.~\eqref{eq:CPstar_definition}, contains the light-flavor and heavy-antiquark magnetic contributions.  The light-quark charge assignments are \(Q_u=2/3\) and \(Q_d=Q_s=-1/3\), while the heavy-antiquark charges are \(Q_{\bar c}=-2/3\) and \(Q_{\bar b}=1/3\).  In the charm sector, this gives \(C_{\bar D^{*0}}=2\tilde a/3-2a\) and \(C_{D^{*-}}=C_{D_s^{*-}}=-\tilde a/3-2a\), whereas in the bottom sector, \(C_{B^{*+}}=2\tilde a/3+a\) and \(C_{B^{*0}}=C_{B_s^{*0}}=-\tilde a/3+a\).

\subsection{\texorpdfstring{\(\pi^0\)- and \(\eta\)-Pole Contributions}{pi0- and eta-Pole Contributions}}

The \(\phi\gamma\gamma\) interaction generated by the chiral anomaly through the Wess--Zumino--Witten functional in Eq.~\eqref{eq:WZW_phigammagamma}, together with the axial \(P^*P^*\phi\) vertex in Eq.~\eqref{eq:LO_HMChPT}, generates the \(t\)-channel exchange of the neutral Goldstone bosons \(\phi=\pi^0,\eta\).  This mechanism contributes only to \(A_5\) and \(A_6\):
\begin{align}
A_{5}^{(\pi^0,\eta)\text{-pole}}
&=\frac{e^2g}{8\pi^2}\,\omega^3
\sum_{\phi=\pi^0,\eta}
\frac{C_{\phi}}{F_\phi^2M_\phi^2},
\notag\\
A_{6}^{(\pi^0,\eta)\text{-pole}}
&=-\frac{e^2g}{8\pi^2}\,\omega^3
\sum_{\phi=\pi^0,\eta}
\frac{C_{\phi}}{F_\phi^2M_\phi^2}.
\label{eq:neutral_pole_forward}
\end{align}
The flavor coefficients \(C_\phi\) depend on the light flavor \(q=u,d,s\) of the external vector meson; their values are collected in Table~\ref{tab:spin1_flavor_coefficients}.

\subsection{\texorpdfstring{\(P^*\phi\)}{P*phi} Loop Contributions}

For the loop contributions, the Goldstone boson is \(\phi=\pi,K\).  We introduce the integration convention
\begin{equation}
\int_\triangle \dd x\,\dd y
\equiv \int_0^1\dd x\int_0^{1-x}\dd y,\qquad
z=1-x-y.
\label{eq:triangle_measure}
\end{equation}
The pion and kaon loops are weighted by the flavor coefficients \(D_\phi^{(c)}\), \(D_\phi^{(d+e)}\), \(D_\phi^{(f)}\), and \(D_\phi^{(g)}\), whose values are given in Table~\ref{tab:spin1_flavor_coefficients}.  A vanishing entry eliminates the corresponding loop contribution.  The auxiliary polynomials entering the two-parameter integrals are
\begin{widetext}
\begin{align}
p_1&=x^2+y^2-x-y+x(9y-2)-2y+1,
&
p_2&=x(7y-1)-y,
\notag\\
p_3&=x+y-1,
&
p_5&=-3x-3y+2,
\notag\\
p_6&=x+y,
&
p_7&=1-2x^2-2y^2+2xy,
\notag\\
p_8&=2x^2+2y^2-2xy-1,
&
p_9&=x(8y-3)-3y+2,
\notag\\
p_{10}&=-2x^2+x-2y^2+y,
&
p_{11}&=3x^2-2x+y(3y-2),
\notag\\
q_1&=q_2=xy z^2,
&
q_5&=xy(x+y-1),
\notag\\
q_6&=-xy(x+y-1),
&
q_9&=xy(-2xy+x+y-1),
\notag\\
q_{10}&=xy\left[x(2y-1)-y+1\right],
&
q_{11}&=xy(-x^2+x-y^2+y),
\notag\\
q_{12}&=xy\left[x^2-x+(y-1)y\right].
\label{eq:forward_polynomials}
\end{align}
\end{widetext}

\begin{table*}[t]
\centering
\renewcommand{\arraystretch}{1.35}
\setlength{\tabcolsep}{5pt}
\caption{Flavor coefficients for the neutral-Goldstone pole and Goldstone-boson loop contributions.  The coefficients \(C_{\pi^0}\) and \(C_\eta\) multiply the corresponding pole terms in Eq.~\eqref{eq:neutral_pole_forward}.  For a fixed external vector meson, \(D_\phi^{(c)}\), \(D_\phi^{(d+e)}\), \(D_\phi^{(f)}\), and \(D_\phi^{(g)}\) multiply the contributions from the corresponding pion and kaon loop topologies.  The loop coefficients follow the flavor-trace convention of Ref.~\cite{Dang:2026bxe}. The light-flavor labels \(q=u,d,s\) correspond, respectively, to \(\bar D^{*0}\), \(D^{*-}\), and \(D_s^{*-}\) in the charm sector and to \(B^{*+}\), \(B^{*0}\), and \(B_s^{*0}\) in the bottom sector.}
\label{tab:spin1_flavor_coefficients}
\begin{tabular}{c|cc|cccc|cccc}
\hline\hline
\multirow{2}{*}{\(q\)}
& \multicolumn{2}{c|}{Neutral-Goldstone pole}
& \multicolumn{4}{c|}{\(\pi\)-loop}
& \multicolumn{4}{c}{\(K\)-loop} \\
\cline{2-3}\cline{4-7}\cline{8-11}
& \(C_{\pi^0}\) & \(C_\eta\)
& \(D_\pi^{(c)}\) & \(D_\pi^{(d+e)}\) & \(D_\pi^{(f)}\)
& \(D_\pi^{(g)}\)
& \(D_K^{(c)}\) & \(D_K^{(d+e)}\) & \(D_K^{(f)}\)
& \(D_K^{(g)}\) \\
\hline
\(u\) & \(1\) & \(1/3\)
      & \(-1\) & \(4\) & \(-4\) & \(1\)
      & \(-1\) & \(4\) & \(-4\) & \(1\) \\
\(d\) & \(-1\) & \(1/3\)
      & \(-1\) & \(4\) & \(-4\) & \(1\)
      & \(0\) & \(0\) & \(0\) & \(0\) \\
\(s\) & \(0\) & \(-2/3\)
      & \(0\) & \(0\) & \(0\) & \(0\)
      & \(-1\) & \(4\) & \(-4\) & \(1\) \\
\hline\hline
\end{tabular}
\end{table*}

For a \(P^*\phi\) intermediate state, the residual-energy argument of the loop functions is \(\delta=0\).  The corresponding amplitudes are
\begin{widetext}
\begin{align}
A_1^{(P^*\phi)}={}&
-\sum_{\phi=\pi,K}\frac{e^2g^2D_\phi^{(c)}}{2F_\phi^2}
\frac23\calJ_0(\omega,0,M_\phi^2)
-\sum_{\phi=\pi,K}\frac{e^2g^2D_\phi^{(d+e)}}{2F_\phi^2}
\int_0^1\dd x\,\frac23\calJ_2'(x\omega,0,M_\phi^2)
\notag\\
&-\sum_{\phi=\pi,K}\frac{e^2g^2D_\phi^{(f)}}{2F_\phi^2}
\int_\triangle\dd x\,\dd y
\left[
\frac{10}{3}\calJ_6''(z\omega,0,M_\phi^2)
-\frac23\omega^2p_1\calJ_2''(z\omega,0,M_\phi^2)
+\frac23\omega^4q_1\calJ_0''(z\omega,0,M_\phi^2)
\right]
\notag\\
&-\sum_{\phi=\pi,K}\frac{e^2g^2D_\phi^{(g)}}{2F_\phi^2}
\int_0^1\dd x\,(d-2)\calJ_2'(0,0,M_\phi^2),
\notag\\
A_2^{(P^*\phi)}={}&
-\sum_{\phi=\pi,K}\frac{e^2g^2D_\phi^{(f)}}{2F_\phi^2}
\int_\triangle\dd x\,\dd y
\left[
\frac23\omega^2p_2\calJ_2''(z\omega,0,M_\phi^2)
-\frac23\omega^4q_2\calJ_0''(z\omega,0,M_\phi^2)
\right],
\notag\\
A_3^{(P^*\phi)}={}&
-\sum_{\phi=\pi,K}\frac{e^2g^2D_\phi^{(c)}}{2F_\phi^2}
\frac12\calG_0(\omega,0,M_\phi^2)
-\sum_{\phi=\pi,K}\frac{e^2g^2D_\phi^{(d+e)}}{2F_\phi^2}
\int_0^1\dd x\,\frac12\calG_2'(x\omega,0,M_\phi^2)
\notag\\
&-\sum_{\phi=\pi,K}\frac{e^2g^2D_\phi^{(f)}}{2F_\phi^2}
\int_\triangle\dd x\,\dd y\,
\frac12\omega^2p_3\calG_2''(z\omega,0,M_\phi^2),
\notag\\
A_4^{(P^*\phi)}={}&
-\sum_{\phi=\pi,K}\frac{e^2g^2D_\phi^{(f)}}{2F_\phi^2}
\int_\triangle\dd x\,\dd y\,
\frac12\omega^2p_3\calG_2''(z\omega,0,M_\phi^2),
\notag\\
A_5^{(P^*\phi)}={}&
-\sum_{\phi=\pi,K}\frac{e^2g^2D_\phi^{(f)}}{2F_\phi^2}
\int_\triangle\dd x\,\dd y
\left[
\frac14\omega^2p_5\calG_2''(z\omega,0,M_\phi^2)
+\frac12\omega^4q_5\calG_0''(z\omega,0,M_\phi^2)
\right],
\notag\\
A_6^{(P^*\phi)}={}&
-\sum_{\phi=\pi,K}\frac{e^2g^2D_\phi^{(f)}}{2F_\phi^2}
\int_\triangle\dd x\,\dd y
\left[
\frac14\omega^2p_6\calG_2''(z\omega,0,M_\phi^2)
+\frac12\omega^4q_6\calG_0''(z\omega,0,M_\phi^2)
\right],
\notag\\
A_7^{(P^*\phi)}={}&
\sum_{\phi=\pi,K}\frac{e^2g^2D_\phi^{(c)}}{2F_\phi^2}
\frac12\calJ_0(\omega,0,M_\phi^2)
+\sum_{\phi=\pi,K}\frac{e^2g^2D_\phi^{(d+e)}}{2F_\phi^2}
\int_0^1\dd x\,\frac12\calJ_2'(x\omega,0,M_\phi^2)
\notag\\
&+\sum_{\phi=\pi,K}\frac{e^2g^2D_\phi^{(f)}}{2F_\phi^2}
\int_\triangle\dd x\,\dd y
\left[
\calJ_6''(z\omega,0,M_\phi^2)
+\frac12\omega^2p_7\calJ_2''(z\omega,0,M_\phi^2)
\right]
\notag\\
&+\sum_{\phi=\pi,K}\frac{e^2g^2D_\phi^{(g)}}{2F_\phi^2}
\int_0^1\dd x\,x(1-x)\omega^2
\calJ_0'(0,0,M_\phi^2),
\notag\\
A_8^{(P^*\phi)}={}&
-\sum_{\phi=\pi,K}\frac{e^2g^2D_\phi^{(f)}}{2F_\phi^2}
\int_\triangle\dd x\,\dd y\,
\frac12\omega^2p_8\calJ_2''(z\omega,0,M_\phi^2)
+\sum_{\phi=\pi,K}\frac{e^2g^2D_\phi^{(g)}}{2F_\phi^2}
\int_0^1\dd x\,x(1-x)\omega^2
\calJ_0'(0,0,M_\phi^2),
\notag\\
A_9^{(P^*\phi)}={}&
-\sum_{\phi=\pi,K}\frac{e^2g^2D_\phi^{(f)}}{2F_\phi^2}
\int_\triangle\dd x\,\dd y
\left[
\frac14\omega^2p_9\calJ_2''(z\omega,0,M_\phi^2)
+\frac12\omega^4q_9\calJ_0''(z\omega,0,M_\phi^2)
\right]
\notag\\
&-\sum_{\phi=\pi,K}\frac{e^2g^2D_\phi^{(g)}}{2F_\phi^2}
\int_0^1\dd x\,x(1-x)\omega^2
\calJ_0'(0,0,M_\phi^2),
\notag\\
A_{10}^{(P^*\phi)}={}&
-\sum_{\phi=\pi,K}\frac{e^2g^2D_\phi^{(f)}}{2F_\phi^2}
\int_\triangle\dd x\,\dd y
\left[
\frac14\omega^2p_{10}\calJ_2''(z\omega,0,M_\phi^2)
+\frac12\omega^4q_{10}\calJ_0''(z\omega,0,M_\phi^2)
\right],
\notag\\
A_{11}^{(P^*\phi)}={}&
-\sum_{\phi=\pi,K}\frac{e^2g^2D_\phi^{(f)}}{2F_\phi^2}
\int_\triangle\dd x\,\dd y
\left[
\frac12\omega^2p_{11}\calJ_2''(z\omega,0,M_\phi^2)
+\frac12\omega^4q_{11}\calJ_0''(z\omega,0,M_\phi^2)
\right]
\notag\\
&+\sum_{\phi=\pi,K}\frac{e^2g^2D_\phi^{(g)}}{2F_\phi^2}
\int_0^1\dd x\,x(1-x)\omega^2
\calJ_0'(0,0,M_\phi^2),
\notag\\
A_{12}^{(P^*\phi)}={}&
-\sum_{\phi=\pi,K}\frac{e^2g^2D_\phi^{(f)}}{2F_\phi^2}
\int_\triangle\dd x\,\dd y\,
\frac12\omega^4q_{12}\calJ_0''(z\omega,0,M_\phi^2).
\label{eq:Astar_forward}
\end{align}
\end{widetext}

\subsection{\texorpdfstring{\(P\phi\)}{Pphi} Loop Contributions}

For a \(P\phi\) intermediate state, the residual-energy argument is channel dependent and equals \(\delta_{\rm ch}\) as defined in Eq.~\eqref{eq:channel_threshold_definitions}. To keep the analytic expressions compact, we denote this argument generically by \(-\Delta\) below. In the numerical evaluation, every occurrence of \(-\Delta\) under the flavor sum is replaced by the corresponding channel-specific \(\delta_{\rm ch}\). The integration measure, auxiliary polynomials, and flavor coefficients are those introduced above. The corresponding amplitudes are
\begin{widetext}
\begin{align}
A_1^{(P\phi)}={}&
-\sum_{\phi=\pi,K}\frac{e^2g^2D_\phi^{(c)}}{2F_\phi^2}
\frac13\calJ_0(\omega,-\Delta,M_\phi^2)
-\sum_{\phi=\pi,K}\frac{e^2g^2D_\phi^{(d+e)}}{2F_\phi^2}
\int_0^1\dd x\,\frac13\calJ_2'(x\omega,-\Delta,M_\phi^2)
\notag\\
&-\sum_{\phi=\pi,K}\frac{e^2g^2D_\phi^{(f)}}{2F_\phi^2}
\int_\triangle\dd x\,\dd y
\left[
\frac53\calJ_6''(z\omega,-\Delta,M_\phi^2)
-\frac13\omega^2p_1\calJ_2''(z\omega,-\Delta,M_\phi^2)
+\frac13\omega^4q_1\calJ_0''(z\omega,-\Delta,M_\phi^2)
\right]
\notag\\
&-\sum_{\phi=\pi,K}\frac{e^2g^2D_\phi^{(g)}}{2F_\phi^2}
\int_0^1\dd x\,\frac{d-2}{2}\calJ_2'(0,-\Delta,M_\phi^2),
\notag\\
A_2^{(P\phi)}={}&
-\sum_{\phi=\pi,K}\frac{e^2g^2D_\phi^{(f)}}{2F_\phi^2}
\int_\triangle\dd x\,\dd y
\left[
\frac13\omega^2p_2\calJ_2''(z\omega,-\Delta,M_\phi^2)
-\frac13\omega^4q_2\calJ_0''(z\omega,-\Delta,M_\phi^2)
\right],
\notag\\
A_3^{(P\phi)}={}&
-\sum_{\phi=\pi,K}\frac{e^2g^2D_\phi^{(c)}}{2F_\phi^2}
\frac12\calG_0(\omega,-\Delta,M_\phi^2)
-\sum_{\phi=\pi,K}\frac{e^2g^2D_\phi^{(d+e)}}{2F_\phi^2}
\int_0^1\dd x\,\frac12\calG_2'(x\omega,-\Delta,M_\phi^2)
\notag\\
&-\sum_{\phi=\pi,K}\frac{e^2g^2D_\phi^{(f)}}{2F_\phi^2}
\int_\triangle\dd x\,\dd y\,
\frac12\omega^2p_3\calG_2''(z\omega,-\Delta,M_\phi^2),
\notag\\
A_4^{(P\phi)}={}&
-\sum_{\phi=\pi,K}\frac{e^2g^2D_\phi^{(f)}}{2F_\phi^2}
\int_\triangle\dd x\,\dd y\,
\frac12\omega^2p_3\calG_2''(z\omega,-\Delta,M_\phi^2),
\notag\\
A_5^{(P\phi)}={}&
-\sum_{\phi=\pi,K}\frac{e^2g^2D_\phi^{(f)}}{2F_\phi^2}
\int_\triangle\dd x\,\dd y
\left[
\frac14\omega^2p_5\calG_2''(z\omega,-\Delta,M_\phi^2)
+\frac12\omega^4q_5\calG_0''(z\omega,-\Delta,M_\phi^2)
\right],
\notag\\
A_6^{(P\phi)}={}&
-\sum_{\phi=\pi,K}\frac{e^2g^2D_\phi^{(f)}}{2F_\phi^2}
\int_\triangle\dd x\,\dd y
\left[
\frac14\omega^2p_6\calG_2''(z\omega,-\Delta,M_\phi^2)
+\frac12\omega^4q_6\calG_0''(z\omega,-\Delta,M_\phi^2)
\right],
\notag\\
A_7^{(P\phi)}={}&
-\sum_{\phi=\pi,K}\frac{e^2g^2D_\phi^{(c)}}{2F_\phi^2}
\frac12\calJ_0(\omega,-\Delta,M_\phi^2)
-\sum_{\phi=\pi,K}\frac{e^2g^2D_\phi^{(d+e)}}{2F_\phi^2}
\int_0^1\dd x\,\frac12\calJ_2'(x\omega,-\Delta,M_\phi^2)
\notag\\
&-\sum_{\phi=\pi,K}\frac{e^2g^2D_\phi^{(f)}}{2F_\phi^2}
\int_\triangle\dd x\,\dd y
\left[
\calJ_6''(z\omega,-\Delta,M_\phi^2)
+\frac12\omega^2p_7\calJ_2''(z\omega,-\Delta,M_\phi^2)
\right]
\notag\\
&-\sum_{\phi=\pi,K}\frac{e^2g^2D_\phi^{(g)}}{2F_\phi^2}
\int_0^1\dd x\,x(1-x)\omega^2
\calJ_0'(0,-\Delta,M_\phi^2),
\notag\\
A_8^{(P\phi)}={}&
\sum_{\phi=\pi,K}\frac{e^2g^2D_\phi^{(f)}}{2F_\phi^2}
\int_\triangle\dd x\,\dd y\,
\frac12\omega^2p_8\calJ_2''(z\omega,-\Delta,M_\phi^2)
-\sum_{\phi=\pi,K}\frac{e^2g^2D_\phi^{(g)}}{2F_\phi^2}
\int_0^1\dd x\,x(1-x)\omega^2
\calJ_0'(0,-\Delta,M_\phi^2),
\notag\\
A_9^{(P\phi)}={}&
\sum_{\phi=\pi,K}\frac{e^2g^2D_\phi^{(f)}}{2F_\phi^2}
\int_\triangle\dd x\,\dd y
\left[
\frac14\omega^2p_9\calJ_2''(z\omega,-\Delta,M_\phi^2)
+\frac12\omega^4q_9\calJ_0''(z\omega,-\Delta,M_\phi^2)
\right]
\notag\\
&+\sum_{\phi=\pi,K}\frac{e^2g^2D_\phi^{(g)}}{2F_\phi^2}
\int_0^1\dd x\,x(1-x)\omega^2
\calJ_0'(0,-\Delta,M_\phi^2),
\notag\\
A_{10}^{(P\phi)}={}&
\sum_{\phi=\pi,K}\frac{e^2g^2D_\phi^{(f)}}{2F_\phi^2}
\int_\triangle\dd x\,\dd y
\left[
\frac14\omega^2p_{10}\calJ_2''(z\omega,-\Delta,M_\phi^2)
+\frac12\omega^4q_{10}\calJ_0''(z\omega,-\Delta,M_\phi^2)
\right],
\notag\\
A_{11}^{(P\phi)}={}&
\sum_{\phi=\pi,K}\frac{e^2g^2D_\phi^{(f)}}{2F_\phi^2}
\int_\triangle\dd x\,\dd y
\left[
\frac12\omega^2p_{11}\calJ_2''(z\omega,-\Delta,M_\phi^2)
+\frac12\omega^4q_{11}\calJ_0''(z\omega,-\Delta,M_\phi^2)
\right]
\notag\\
&-\sum_{\phi=\pi,K}\frac{e^2g^2D_\phi^{(g)}}{2F_\phi^2}
\int_0^1\dd x\,x(1-x)\omega^2
\calJ_0'(0,-\Delta,M_\phi^2),
\notag\\
A_{12}^{(P\phi)}={}&
+\sum_{\phi=\pi,K}\frac{e^2g^2D_\phi^{(f)}}{2F_\phi^2}
\int_\triangle\dd x\,\dd y\,
\frac12\omega^4q_{12}\calJ_0''(z\omega,-\Delta,M_\phi^2).
\label{eq:AP_forward}
\end{align}
\end{widetext}

Equations~\eqref{eq:born_b2_forward}, \eqref{eq:neutral_pole_forward}, \eqref{eq:Astar_forward}, and \eqref{eq:AP_forward}, together with Table~\ref{tab:spin1_flavor_coefficients}, provide all contributions required to extract the polarizabilities through Eq.~\eqref{eq:Pi_definition}.

\end{appendix}

\bibliography{references}

\end{document}